 \documentclass[preprint,5p,times,twocolumn]{elsarticle}
\usepackage{hyperref}
\usepackage{xcolor}
\usepackage[T1]{fontenc}
\usepackage[utf8]{inputenc}
\hypersetup{
    colorlinks=true,
    linkcolor=red,
    citecolor=red,
    urlcolor=red
}
\usepackage{graphicx,amsmath} % Required for inserting images

\usepackage{caption}
\usepackage{subfig}
\usepackage{lipsum}

\usepackage{cleveref}
\usepackage{cuted}
\usepackage{float}
\usepackage{xr}

\def\vn{\mathbf{n}}
\def\vx{\mathbf{x}}
\usepackage{amssymb}
\usepackage{lipsum}

\journal{Carbon}

\begin{document}

\begin{frontmatter}



    \title{Nanoconfined Water Phase Transitions in Infinite Graphene Slits: Molecular Dynamics Simulations and Mean-Field Insights}

\author[1]{Felipe Hawthorne}
\author[1]{Virgília M. S. Neta}
\author[1]{José A. Freire }
\author[1]{Cristiano F. Woellner }

\affiliation[1]{organization={Physics Department, Federal University of Paraná, UFPR},%Department and Organization
            city={Curitiba},
            state={PR},
            postcode={81531-990}, 
            country={Brazil}}

%, UFPR, Curitiba, PR, 81531-980, Brazil
\begin{abstract}

%Experimental and computational results have recently shown that the nanoconfinement of water is able to promote a complex phase behavior at much lower pressures and temperatures than usually expected. Nanoconfinement, when coupled with adsorption, has also been documented to promote a much higher uptake due to the occuring condensation at the isotherm curve's onset, a direct consequence of the more easily formed Hydrogen bond networks. We propose a two-way approach in order to investigate this phenomena in water confined in infinite graphene slits. Our Molecular Dynamics results show that hysteresis is observed for all temperatures and layer sizes investigated, predicting different phase behaviors dependent on the formation of hydrogen bond networks, directly associated with a higher water uptake. Our proposed mean-field approach based on the Ono-Kondo model was also able to adequately parameterize the isotherm curves, as well as provide a more complete description (at mean-field level) of adsorption in nanoconfinement, irrespective of materials. 

Recent experimental and computational studies have demonstrated that nanoconfinement profoundly alters the phase behavior of water, facilitating complex phase transitions at pressures and temperatures far lower than typically observed in bulk systems. When combined with adsorption, nanoconfinement substantially enhances water uptake, primarily due to condensation occurring at the onset of the isotherm curve—a phenomenon intimately related to the facilitated formation of hydrogen bond networks. In this study, we adopt a dual approach to investigate water confined within infinite graphene slits. Our Molecular Dynamics simulations reveal hysteresis across all investigated temperatures. Unlike in finite slits, where hysteresis arises due to surface tension effects at the edges, in the case of infinite slits, the hysteresis is the result of a genuine phase transition at the nanoscale. We analyze the spatial and orientational arrangements of the water molecules, demonstrating how the graphene surface promotes the formation of a hydrogen bond network in the adjacent water layers. The remarkably low pressure required for water uptake in this nano-environment is explained at the mean-field level using a simple interacting lattice model. This is attributed to the exponential dependence of the critical pressure on the adsorbate-adsorbent interaction.

\end{abstract}




\end{frontmatter}

%\tableofcontents



\section{Introduction}
\label{sec:intro}

%The process of adsorption has been shown to be greatly relevant to many different problems, ranging from the production of gas sensors, to filtration of liquids of providing a pathway to understading the complex interactions that rule certain phenomena \cite{Ma2020,brunauer1938adsorption, DABROWSKI2001135,toth2002adsorption,suzuki1990adsorption,ayawei2017modelling,ruthven1984principles,faust2013adsorption}. The process of adsorption of water, although a relatively old problem \cite{BDDT_1940}, be it with the use of Metal-Organic Frameworks (MOFs) \cite{Fei2022, Furukawa2014} or Carbons, such as Graphene, Graphene-Oxide or Activated Carbon \cite{Alcaniz-Monge2001, DO2000767,lian2018extraordinary} or novel 2D materials in general \cite{sacchi}, has recently become one the major focuses of investigation in the realm of condensed matter. Present in many different problems, such as pollutant removal \cite{El-Baz,RATHI2021116995}, improvement of catalysis in chemical processes \cite{Ukhurebor2024, NUPEARACHCHI2017206} and even on the purification of wastewater for pharmaceutical applications \cite{Kryuchkova2021}, the question that arises it not why this process is so important, but whether through what methods and types of adsorbents can this process be fully optimized. 

The process of adsorption has been shown to be highly relevant to a wide range of problems, from the production of gas sensors and liquid filtration to providing insights into the complex interactions governing certain phenomena such as phase transitions and nanoconfinement~\cite{Ma2020,DABROWSKI2001135,toth2002adsorption,Kapil2022,suzuki1990adsorption,ayawei2017modelling,ruthven1984principles,faust2013adsorption,BDDT_1940}. Although the adsorption of water is a relatively old problem \cite{Kanagy1950,MAHAJAN1971308,OVERLOOP1993179,Alcaniz-Monge2001,ma2011adsorption}, recent advancements, particularly with the use of Metal-Organic Frameworks (MOFs) \cite{Fei2022, Furukawa2014,Hastings2024}, carbon based materials such as graphene, graphene oxide, or activated carbon \cite{Alcaniz-Monge2001,DO2000767,lian2018extraordinary,Gkika2022}, and other novel 2D materials \cite{sacchi}, have brought it to the forefront of condensed matter research once again. This process plays a crucial role in various applications, including pollutant removal \cite{El-Baz,RATHI2021116995,NUPEARACHCHI2017206}, enhancement of catalysis in chemical processes \cite{Ukhurebor2024,suzuki1990adsorption}, and even wastewater purification for pharmaceutical purposes \cite{Kryuchkova2021,WONG2019103261,w13020215}. The question is not whether adsorption is important, but rather how this process can be fully optimized and understood through various methods—computational, theoretical, and experimental alike.

%Graphene, a novel carbon allotrope with $sp^2$ hybridization derivative of graphite~\cite{Geim_2009}, and graphene-based materials have been shown to be a great candidate for this purpose, specially due to the emergent interactions when the adsorbate in question is water~\cite{Yang2020,NUPEARACHCHI2017206,Hamada_2012,Algara-Siller2015,WANG2021102360}. Furthermore, the behavior of the latter, the water molecules, have also been observed to deviate from the expected behavior in the adsorption process \cite{Canivet2014, MAHAJAN1971308, Kolle2021}. The presence of hydrogen atoms cause formation of hydrogen bonds, and consequentely, hydrogen bonds networks, specially in confined adsorption \cite{Yang2020,Cai2021,Alloush2024}. Contrary to gas adsorption, such as $CO_2$, where the formation of multi-layer is rarely observed due to the weak interaction between adsorbate molecules \cite{CINKE2003761,Gargiulo2014, YuChengHuan}, the emergence of bulk structures of water during adsorption has been well documented, as well as investigated through many different approaches and materials~\cite{SARKISOV2017127,jiamicro,Woellner_2017,Woellner_2018,ma2011adsorption,liu2017water}

Graphene, along with graphene-based materials, has shown great potential for this purpose \cite{Yang2020,NUPEARACHCHI2017206,Hamada_2012,Algara-Siller2015,WANG2021102360}. Furthermore, the behavior of water (and other hydrogen bearing) molecules during the adsorption process has been observed to deviate from expected norms \cite{Canivet2014, MAHAJAN1971308, Kolle2021}, especially in confined environments, largely due to the formation of Hydrogen bond networks~\cite{Yang2020,Cai2021,Alloush2024}. In contrast to gas adsorption, such as $\rm{CO}_2$, where the formation of multilayers is rarely observed due to the weak interactions between adsorbate molecules \cite{CINKE2003761,Gargiulo2014, YuChengHuan}, the emergence of bulk water structures during adsorption has been well documented and investigated through various approaches and materials \cite{SARKISOV2017127,jiamicro,Woellner_2018,ma2011adsorption,liu2017water}.

The purpose of this paper is to make two distinct contributions to this ever-relevant subject. First, we propose a Molecular Dynamics (MD) approach capable of capturing the full hysteresis cycle of isotherms in the adsorption process. This approach demonstrates that investigating a broader range of temperatures and multilayer structures can provide significant insights into the complex mechanisms and interactions at play, while also providing a robust argument on how confinement can lead to the formation of macroscopic hydrogen bond networks previously observed only at much higher pressures \cite{Kapil2022}. Second, we introduce a mean-field lattice model based on the Ono-Kondo \cite{OnoKondo} model, which explains the hysteresis as a natural consequence of the water condensation under nanonfinement. This model offers a reliable method for determining isotherm characteristic, such as the Henry constant and surface attraction energy, by accounting for complex intrinsic interactions using only two free parameters.

\begin{figure*}[tp]
\centering
\includegraphics[width=\textwidth]{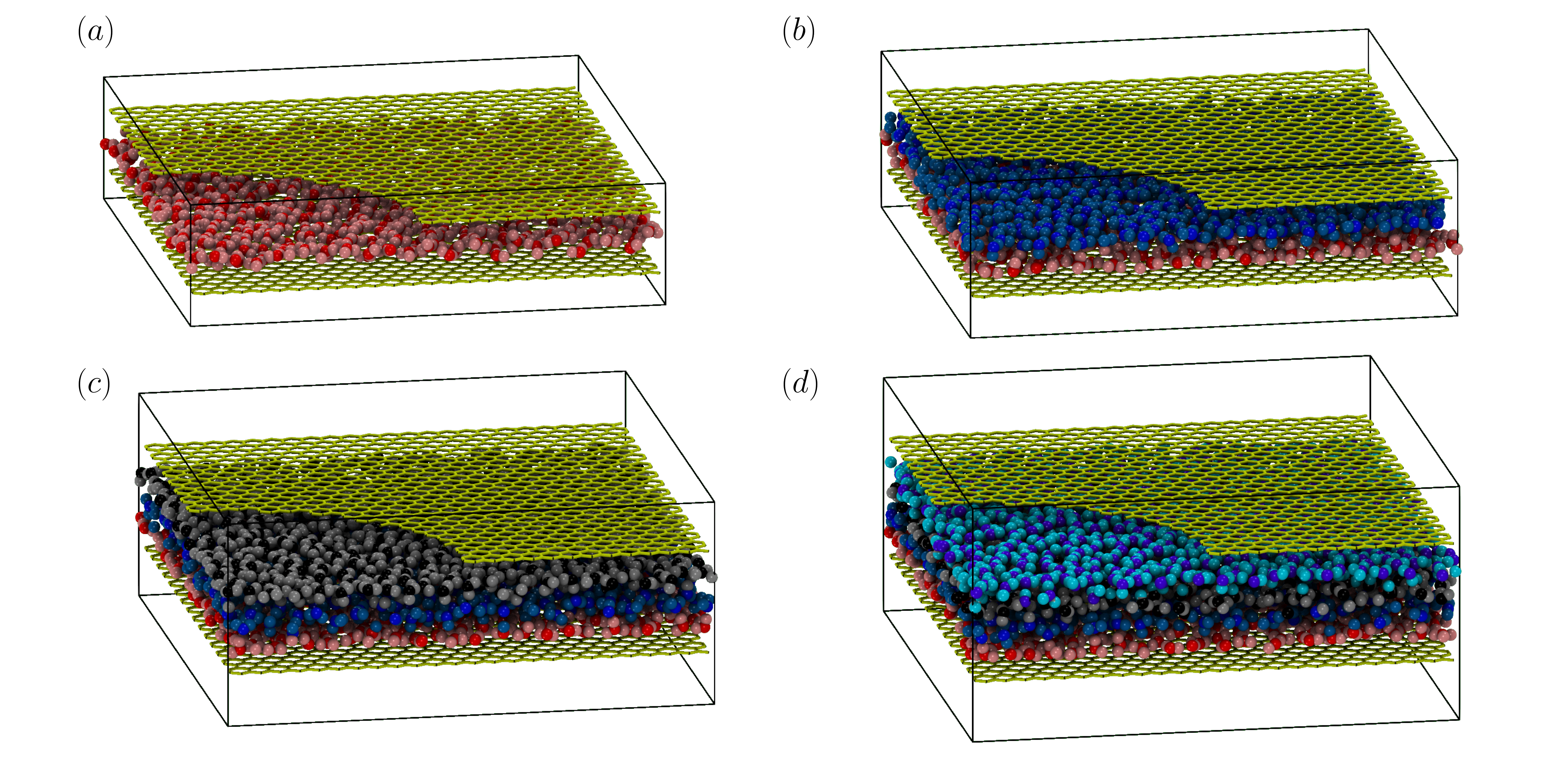}
    \caption{Snapshots of the different configurations investigated in this work at the adsorption saturation limit, as detailed in Section \ref{methods}. Panels $(a)$, $(b)$, $(c)$, and $(d)$ correspond to the structures for the monolayer, bilayer, trilayer, and quadrilayer configurations, respectively. For clarity, in this illustration each layer of water molecules is color-coded (color online) according to its respective layer: red, blue, black, and cyan represent the first, second, third, and fourth layers, respectively. The black lines overlaid on the simulation box in all four panels indicate the presence of periodic boundary conditions in all three directions ($x$, $y$, and $z$). The upper graphene sheets are partially omitted to increase visualization of the adsorbate. }
    \label{figure1}
\end{figure*}

%This paper is organized as follows. In Sec.~\ref{methods}, the methodology implemented in the Molecular Dynamics portion of this work is explained, as well as a brief review of the literature behind the mean-field model presented here. The results obtained via both approaches are then discussed in Sec.~\ref{results}, were we choose to first lay out the Molecular Dynamics results, followed by the mean-field ones, and, finally, a comparison between approaches. Finally, conclusions are drawn in Sec.~\ref{conclusion} and more technical information, as well as links to the results not here included can be found in \ref{ap1} and \ref{ap2}, respectively. 
This paper is organized as follows. In Sec.\ref{methods}, we explain the methodology implemented in the Molecular Dynamics portion of this work and provide a brief review of the literature supporting the mean-field model presented here. The results obtained from both approaches are then discussed in Sec.\ref{results}, where we first present the Molecular Dynamics results, followed by the mean-field results, and conclude with a comparison between the two approaches. Finally, conclusions are drawn in Sec.~\ref{conclusion}, and additional technical details can be found in the Appendix and the supplementary material.

\section{Methods}
\label{methods}
Given that this work presents two distinct approaches to investigating the proposed dynamics, the methods section is divided accordingly.

\subsection{Mean-Field Model}

We considered a lattice model for the adsorption of a gas in an infinite planar slit. This model, originally described by Ono and Kondo \cite{OnoKondo}, extends the well-known BDDT model \cite{BDDT_1940} by incorporating lateral interactions between gas molecules. Donohue and Aranovich \cite{Donogue_1998} later used this model to describe capillary effects in finite planar slits.
\begin{figure}[h!]
\centering
%{\includegraphics[width=0.3\textwidth]{modelo.png}}
\centering{\includegraphics[width=0.35\textwidth]{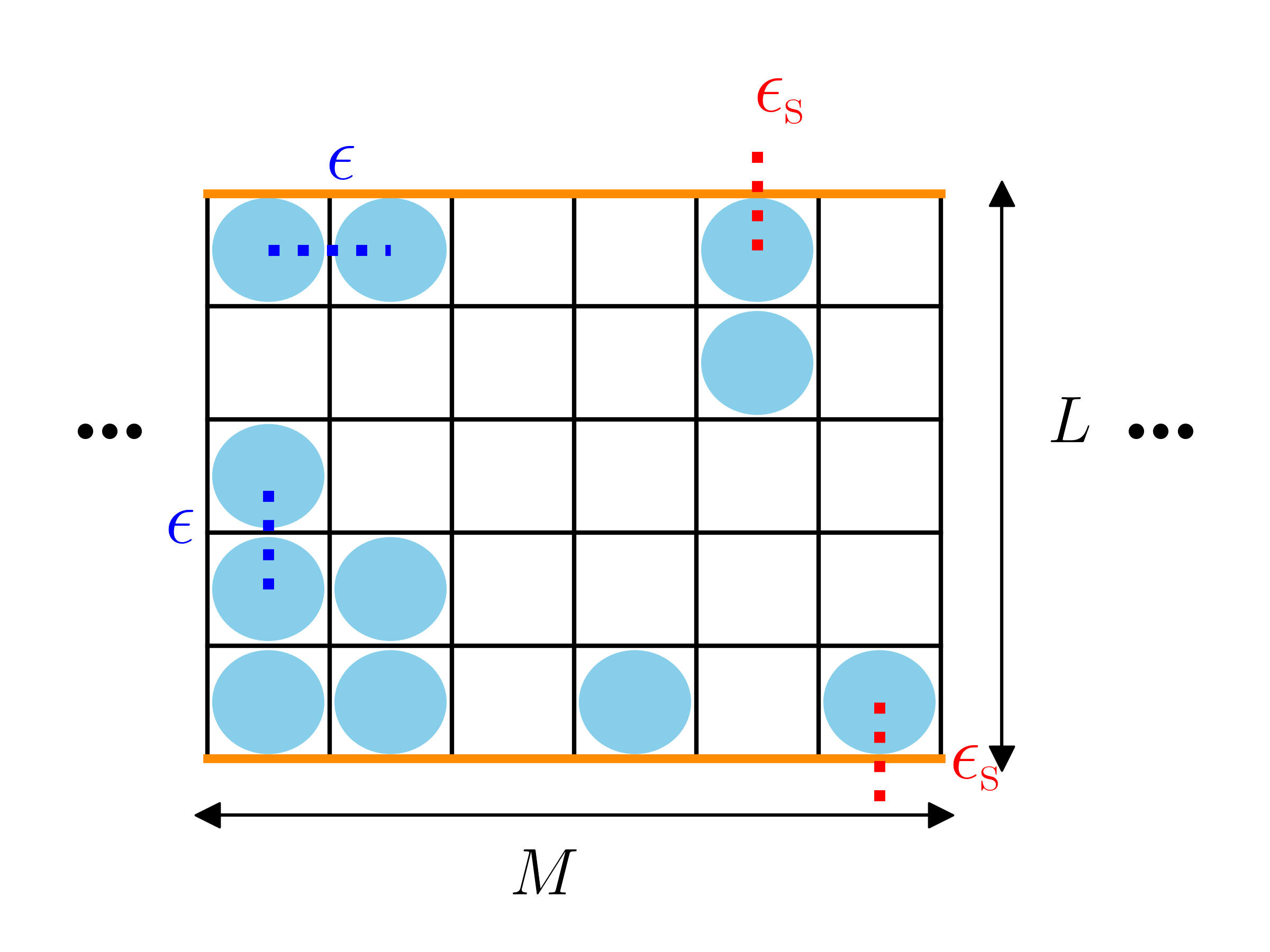}}
\caption{The lattice model for the infinite slit. Lateral periodic boundary conditions were used.}\label{fig:modelo}
\end{figure}

The model, illustrated in Fig. \ref{fig:modelo}, includes both a surface interaction, $\epsilon_s$, and a bulk interaction, $\epsilon$. The system consists of $M$ sites arranged in a square (with lateral periodic boundary conditions) stacked in $L$ layers.

\subsection{Molecular Dynamics}\label{methods:md}

%\FH{The Molecular Dynamics (MD) simulations were performed using LAMMPS~\cite{LAMMPS}. Figure~\ref{figure1} shows the square graphene sheets used in all simulations, each with equal length and width of $w = 60~\AA$ and $N_{\rm C} = 1.5 \times 10^3$ carbon atoms. Four structures, each with different interplanar distances, were investigated. The distance between the graphene sheets of each configuration, $L_i$, where $i = 1,2,3,4$, were chosen to accommodate specific numbers of $H_2O$ layers, \textit{i.e.}, $L_1 = 8 \AA$ for a single (mono) layer, $L_2 = 10 \AA$ for the bilayer, $L_3 = 13\AA$ for the trilayer, and $L_4 = 15 \AA$ for the quadrilayer. The value $L_1$ was chosen based on optimal uptake of water on graphene, as seen in Ref.~\cite{galvao2016}, while subsequent layers $(L_2, L_3, L_4)$ were defined based on the water molecule length and volume. The dimensions of the simulation box are dependent on the structure being investigated, having the same width as the graphene sheet, $60 \AA$, and height $h_i = 2 L_i$. Periodic boundary conditions were imposed on the $x$ and $y$ axis to simulate an infinite slit, while the periodic boundaries on the $z$-axis were utilized so to optimize the data sampling throughout the simulations.}
    
Molecular Dynamics (MD) simulations were performed using LAMMPS~\cite{LAMMPS}. Figure~\ref{figure1} shows the two square graphene sheets used in all simulations, each with a side length of $60~\AA$ and consisting of $1500$ carbon atoms. Four different interplanar distances were investigated. The distances between the graphene sheets in each configuration, denoted as 
$L_i$ (where $i=1,2,3,4$) were selected to accommodate specific numbers of $\rm{H}_2\rm{O}$ layers. Specifically, \( L_1 = 8~\AA \) for a single (mono) layer, \( L_2 = 10~\AA \) for the bilayer, \( L_3 = 13~\AA \) for the trilayer, and \( L_4 = 15 ~\AA \) for the quadrilayer. The value \( L_1 \) was selected based on optimal uptake of water on graphene, as reported in Ref.~\cite{galvao2016}, while subsequent layers \( (L_2, L_3, L_4) \) were based on the water molecule length and volume. Periodic boundary conditions were imposed on the \( x \) and \( y \) axes to simulate an infinite slit, while periodic boundaries on the \( z \)-axis are utilized to optimize data sampling throughout the simulations.

Each initially empty structure is brought into contact with a virtual water reservoir at a fixed temperature. The pressure (and consequently, the chemical potential) of the reservoir was modified in regular increments, and after each change, the system was allowed to evolve for a fixed number of Monte Carlo steps. We analyzed the system's hysteresis behavior during filling and emptying cycles, under different rates of pressure change. A brief discussion regarding the correspondence between Monte Carlo steps, real time, and the values used in this work can be found in Sec.~\ref{time}.

%\FH{For a given set of temperatures and system size, the pressure is initially set to (close to, due to divergence at $p =0$) zero, and is steadily increased throughout the simulation, ending when the saturation limit is reached through adsorption. Afterwards, the pressure is then reduced until the system is then empty again. Throughout this paper, and from here onward, we refer to these two different paths as \textit{forward} and \textit{backward} paths, respectively. Each time an increase (or decrease) in pressure is made, the system is allowed to evolve up to a time of $6 \times 10^5$ Monte Carlo steps, $\tau_{MC}$.The resulting state is then utilized as a the initial condition for the following pressure value. A brief discussion about the correspondence of Monte Carlo steps, real time and the values utilized in this work can be found in Sec  This approach is motivated by the strong dependence on initial conditions and the general instability observed at the onset of the adsorption curve through the loading parameter, which indicates discontinuous-like behavior. This observation is consistent with the hysteresis predicted in the literature on both statistical mechanics and gas adsorption theory~\cite{Donogue_1998, Woellner_2018,berg2004}.}

The graphene sheets were immobile throughout the simulations, and their interaction with water molecules occurs solely through Lennard-Jones interactions. The water molecules were modeled using the rigid extended simple point charge model (SPC/E)~\cite{Berendsen1987}, with partial charges $q_O = -0.846e$, $q_H = |q_O|/2$ and Lennard-Jones cutoff atom distance $\sigma = 3.166~\AA$. The van der Waals interactions were truncated at $14~\AA$ and added Coulombic long-range interactions were calculated via the standard Ewald summation method, with precision of $10^{-5}$~\cite{Ewald,ewald2, wolf}. The full set of parameters used in the Lennard-Jones potential (energy, $\epsilon$, and minimum distance, $\sigma$) can be found in Table \ref{tab:UFF}.

\begin{figure*}[!h]
    \begin{center}
   %\hspace{-1cm}
    \includegraphics[width=0.42\textwidth]{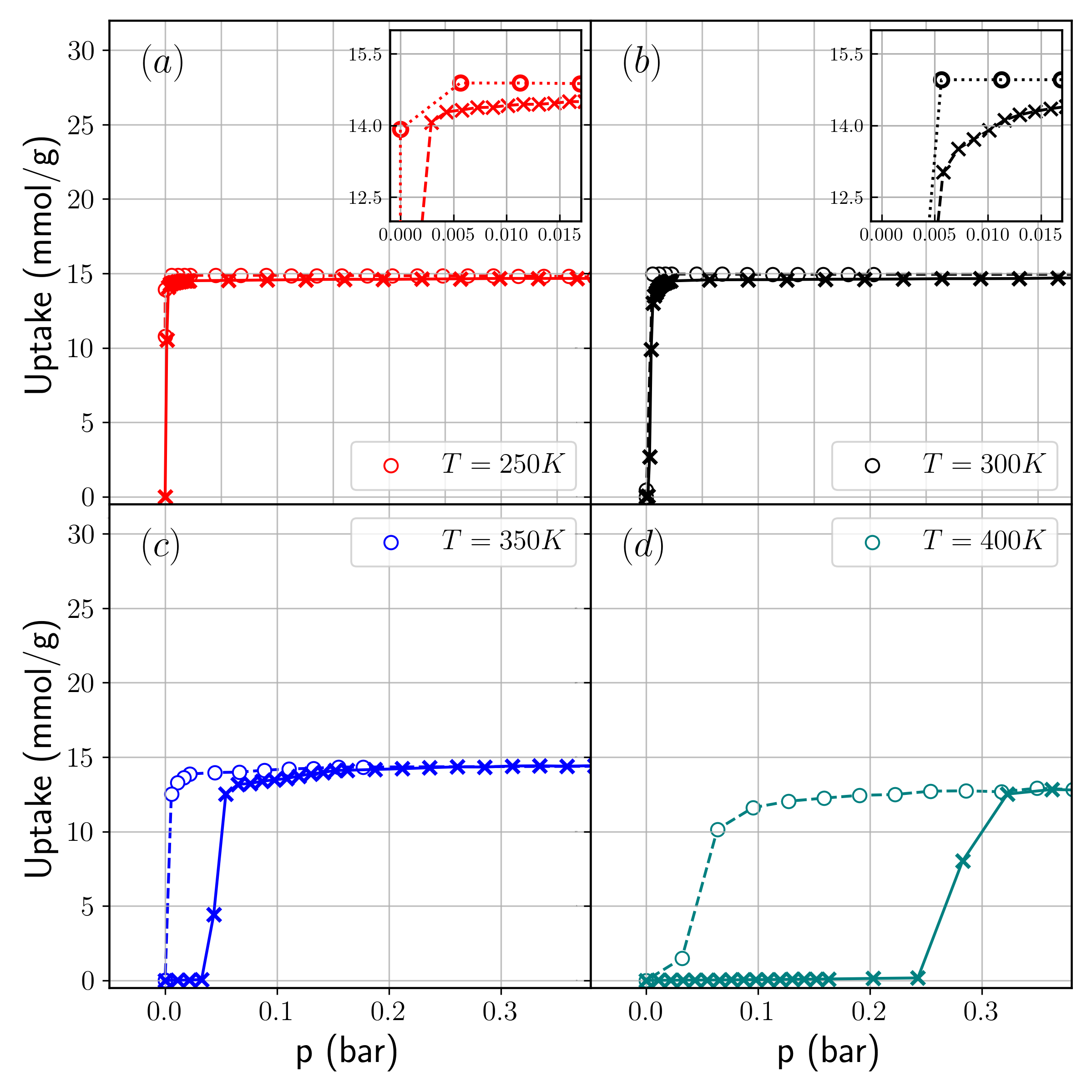}
    \includegraphics[width=0.42\textwidth]{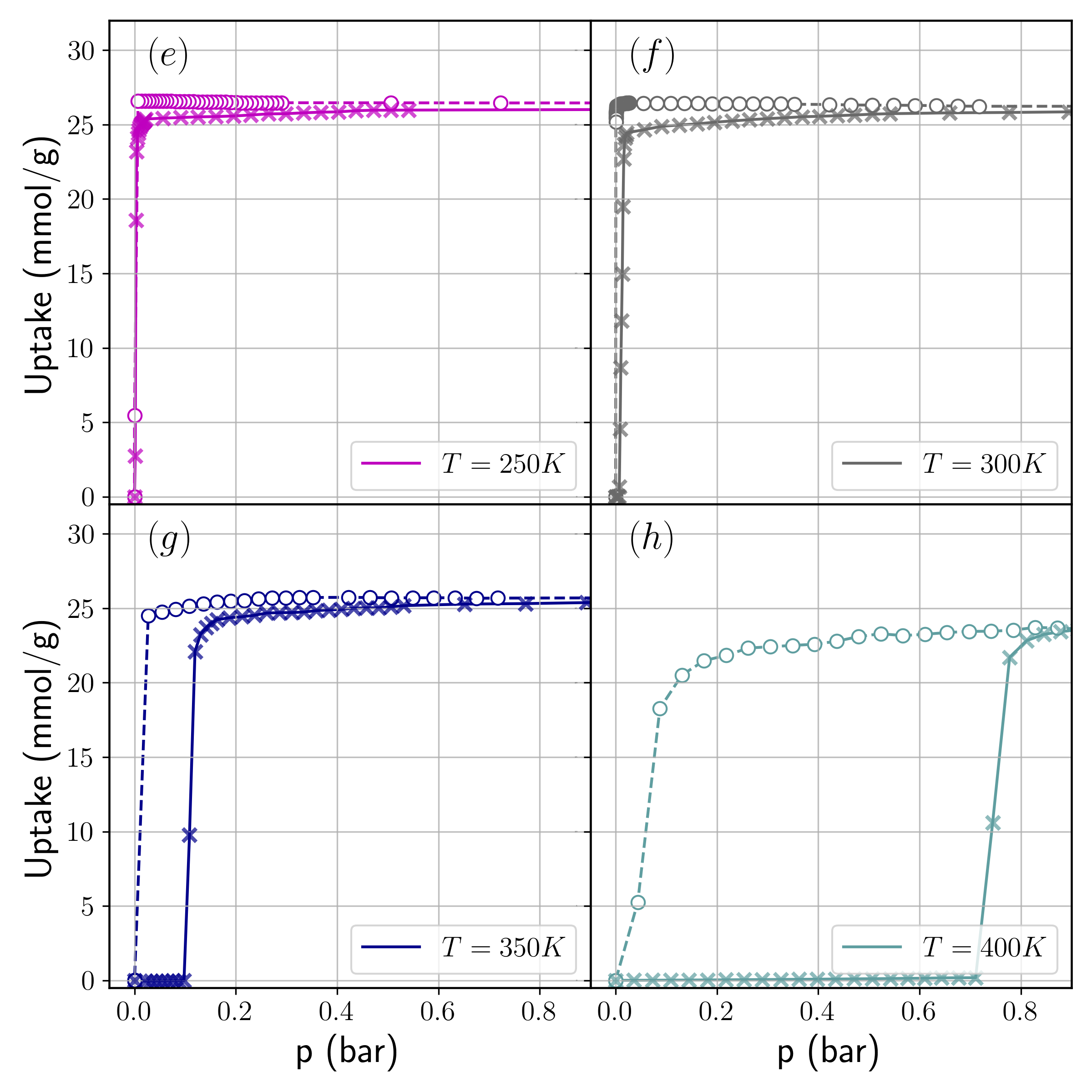}
    \caption{Isotherms for the monolayer, $L_1 = 8~\AA$ at four different temperatures, panels (a) to (d), and  isotherms for the bilayer, $L_2 = 10~\AA$ at the same four temperatures, panels (e) to (h). In all panels, the cross-shaped points indicate the forward path (increasing pressure), while the circled points represent the backward path (decreasing pressure) down to $p = 10^{-10}{\rm bar}$. The insets in panels (a) and (b) provide a zoomed view of the low-pressure region. In all cases the pressure change for the forward path uptake rate was $\kappa_{\rm f} = 4.83 \times 10^{-9}$ bar/MCs, whereas for the backward path, $\kappa_{\rm b} = -7.03 \times 10^{-8}$ bar/MCs.}
    \label{fig:hyst-l8}
    \end{center}
\end{figure*}

%Lorentz~\cite{lorentz1881} and Berthelot ~\cite{berthelot1898} mixing rules were employed to calculate the Lennard-Jones energy, $\epsilon$, and minimum distance, $\sigma$, respectively. 

%The full set of parameters utilized for the LJ interactions is provided in Table \ref{tab:UFF}.

\begin{table}[h!]
    \centering
    \caption{Force field parameters for Lennard-Jones obtained from Refs.~\cite{UFF,Berendsen1987,spce2,Cao2018}}\label{tab:UFF}
    \begin{tabular}{c|ccc}
    \hline \hline
         & $\epsilon $~(kcal/mol) & $\sigma$~($\AA$) &  \\
        \hline
        $H - H$ & $ 0.0000$ & $0.0000$ \\
        $O - O$ & $ 0.1554$ & $3.166$ \\
        $O - H$ & $0.0000$ & $0.0000$\\
        $C - O$ & $0.1284$ & $3.2404$  \\
        $C - H$ & $0.0470$ & $3.0250$ \\
        $C - C$ & $0.1050$ & $ 3.4308$ \\
        \hline \hline
    \end{tabular}
\end{table}

%The Grand Canonical Monte Carlo (GCMC) method is implemented to simulate water adsorption using the GCMC LAMMPS package~\cite{LAMMPS, GCMC1, GCMC2}. The ideal chemical potential \( \mu \) is inferred via the fugacity coefficient \( \phi \), calculated as in Ref.~\cite{GCMC1}, as the pressure changes during the simulation. A temperature damping parameter of \( 100~{\rm fs} \) is used for the thermostat, with a timestep \( \mathrm{d}t = 1~{\rm fs} \), standard for LAMMPS' \textit{real} units style. The GCMC attempts to exchange a water molecule every 100 steps, with 100 attempts made each time. No additional Monte Carlo steps are performed. For each interlayer distance, GCMC simulations are conducted at temperatures \( T \in \left\{250K, 300K, 350K, 400K\right\} \). These temperatures are chosen to facilitate further investigation into water-graphene interactions beyond the room-temperature range commonly observed in the literature, taking advantage of the temperature range covered by the SPC/E model. It is noteworthy that studies of water (as the adsorbate) near or below the freezing point have been conducted in previous works, as seen in Refs.~\cite{Zhang2018, ENDO2012409, OVERLOOP1993179}, though with different materials as adsorbents. The graphene structures are modeled using the Topotools plugin from VMD~\cite{VMD}, and the water molecule using the Avogadro API~\cite{hanwell2012avogadro} and the TrAPPE Database~\cite{trappe, trappe2}.

The Grand Canonical Monte Carlo (GCMC) method was used to set the system's
pressure, which in turn determines the chemical potential
with the aid of the fugacity coefficient, calculated as
in Ref.~\cite{GCMC1},
\begin{equation}
    \mu = k T \log \left(\frac{\phi P \Lambda^3}{k T}\right), 
    \label{eq:mulammps}
\end{equation}
where $P$ refers to the pressure in units of Pa, and
\begin{equation}
    \Lambda = \sqrt{\frac{h^2}{2\pi m k T}},
\end{equation}
is the usual thermal de Broglie wavelength, with $h$ being the Planck constant and $m$ the mass of the adsorbate molecule. In the GCMC method, an attempt to exchange a water molecule with a virtual reservoir was made every 100 steps, with 100 attempts made per exchange cycle.

A temperature damping parameter of $100\times dt$ was used for the thermostat,
with a timestep $dt = 1$~fs. For each interlayer distance, simulations were conducted at temperatures $250$, $300$, $350$, and $400$~K. These temperature values span the range typically explored using the SPC/E water model. It is noteworthy that studies of water near or below the freezing point have been conducted in previous works~\cite{Zhang2018, ENDO2012409, OVERLOOP1993179}, though with different materials as adsorbents. The graphene structures were modeled using the VMD plugin topotools~\cite{VMD}, and the water molecule using the Avogadro API~\cite{hanwell2012avogadro} and the TrAPPE Database~\cite{trappe, trappe2}.

\section{Results and discussion}
\label{results}

\subsection{Temperature effect on the isotherms and the Hydrogen bond network}
\label{results:md}

We begin by examining the results for the monolayer system ($L_1 = 8~\AA$). Our results, see Figs.~\ref{fig:hyst-l8}(a-d), show behaviors consistent with previous studies on the temperature effects of monolayer water adsorption with various adsorbents, e.g., increasing temperature reduces adsorbate maximum uptake (clearly visible at $400K$) ~\cite{Fei2022, Hastings2024, Kanagy1950, Yang2020}. This effect was also manifested in the increase of the adsorption onset pressure (on the forward path) and offset pressure (on the backward path).

%\FH{As shown in Fig.~\ref{fig:layers}, the onset of adsorption occurs at a lower pressure for the monolayer compared to the bilayer, trilayer, and beyond.}

Hysteretic behavior was observed at all temperatures. Since lateral periodic boundary conditions were used, the hysteresis indicates a phase transition rather than the capillary effects typically seen in finite systems~\cite{Donogue_1998}. From an application perspective, it is noteworthy that the maximum water uptake per gram of adsorbate surpasses that of MOFs~\cite{zhangzhu}.

%However, it is noteworthy that the maximum uptake observed for $T\leq 300 K$ not only rivals that seen in some MOFs \cite{zhangzhu}, suggesting that confined adsorption in graphene may outperform other materials, but also that, for this temperature range, the reduction in uptake during the backward path is far more persistent compared to the behavior observed at $T>300$, indicating a change in the transition behavior in this range.

%The aforementioned behavior of water maintaining a high uptake at lower pressures in the backwards path can be attributed to a combination of three different elements that mutually interact to this effect: confinement, hydrogen bonds and the anomalous behavior of water at low temperatures. To further explain this result, an hydrogen-bond analysis was performed at the uptake maximum for the monolayer and all four temperatures, and is presented in Fig.~\ref{fig:hbonds-l8}. Said hydrogen bonds were identified as in Ref.~\cite{Sudheer2017}, defining the cutoff radius and angle as $r_c = 3\AA$ and $\theta_c =\pi $, respectively.

\begin{strip}
%\begin{figure}[h!]
\begin{center}
\includegraphics[width=0.45\textwidth]{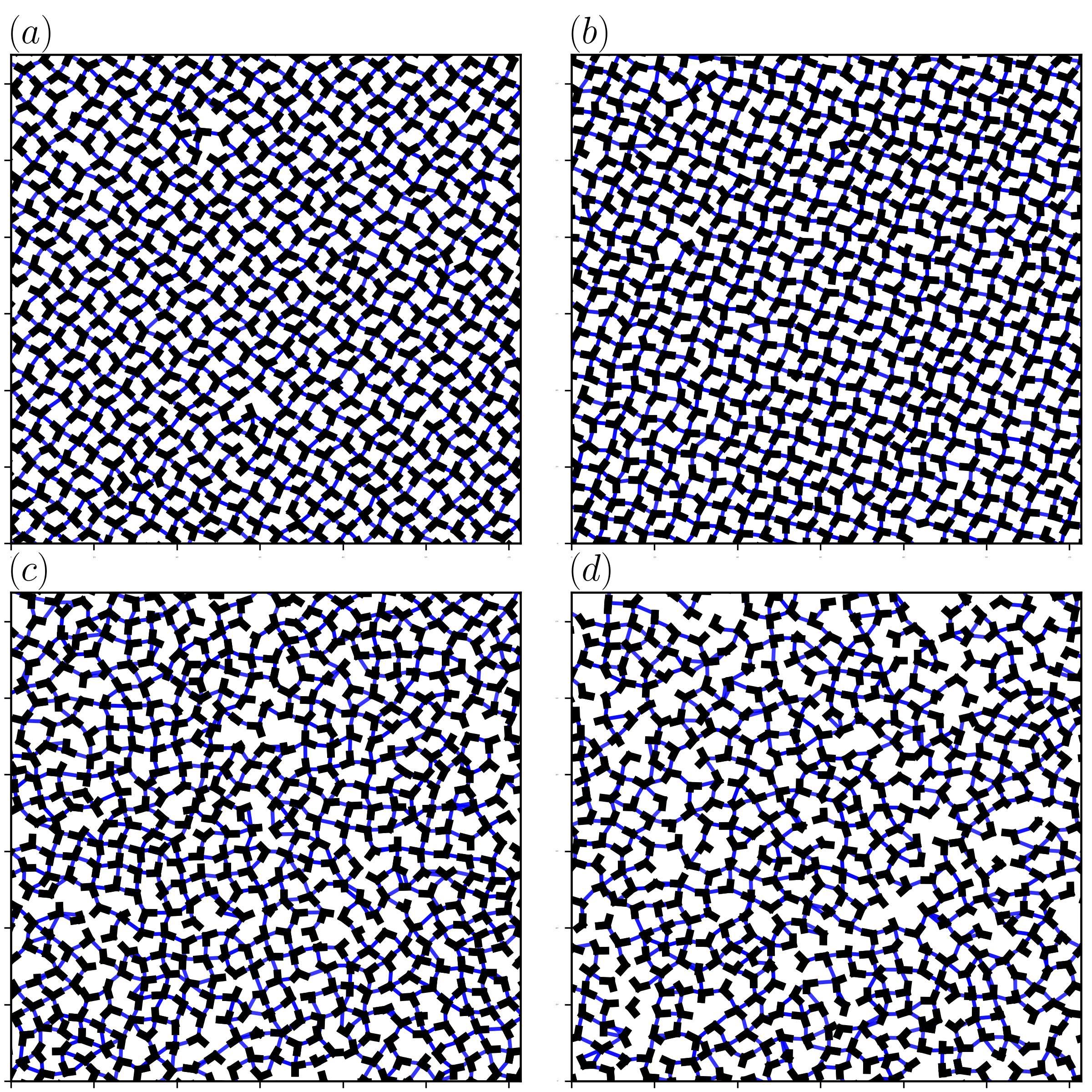}
\includegraphics[width=0.45\textwidth]{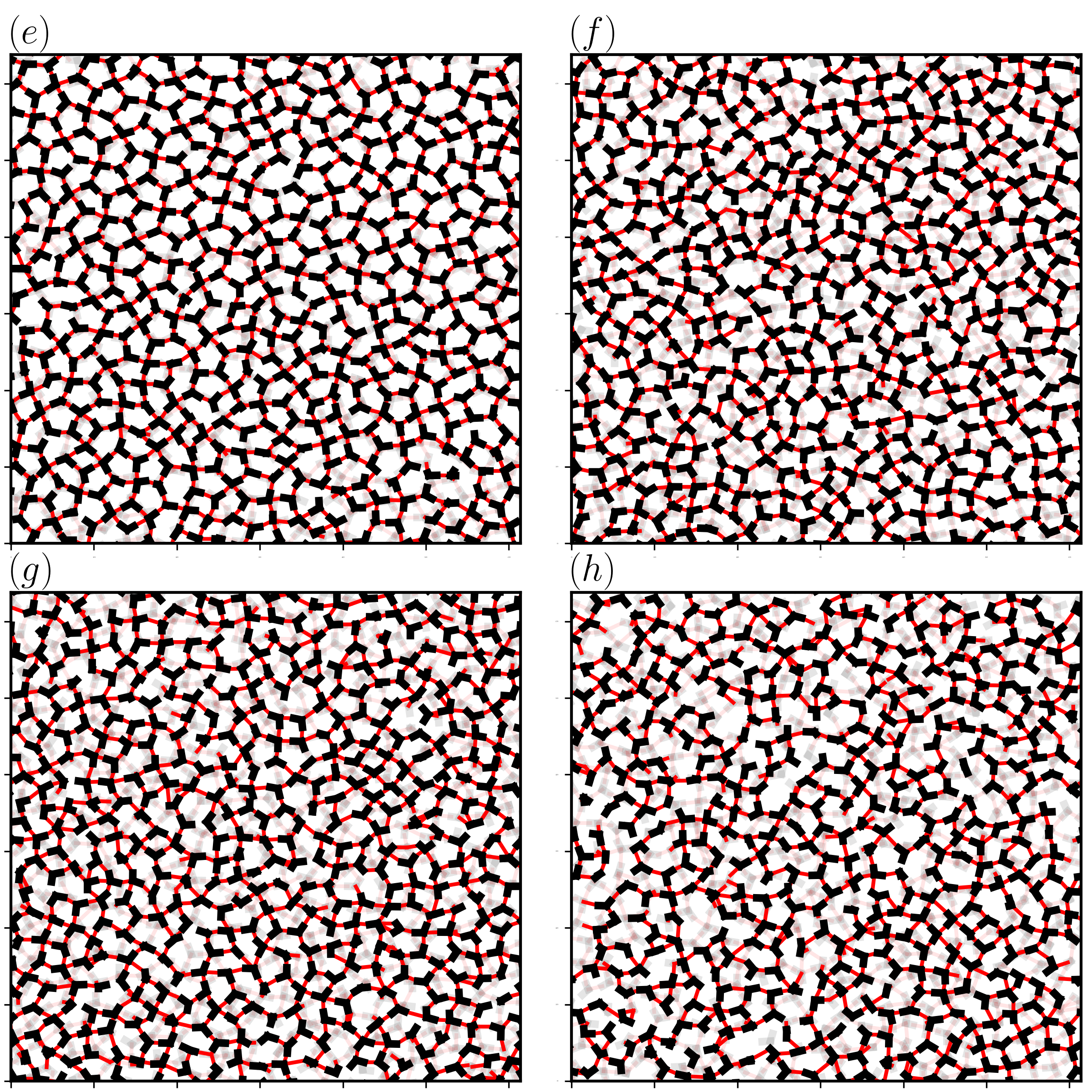}
\end{center}
\captionof{figure}{Illustration of the hydrogen bonds identified in the monolayer using the parameters mentioned in the text and $(a)$ $T = 250$~K, $(b)$ $T = 300$~K,$(c)$ $T = 350$~K and $(d)$ $T = 400$~K, and in the bilayer system for panels $(e)$, $(f)$, $(g)$ and $(h)$ at the same temperatures. Water molecules are represented as black corner-bracket symbols and the Hydrogen bonds are pictured as blue lines for the monolayer and red lines for the bilayer.}
\label{fig:hbonds-l8}
%\end{figure}
\end{strip}

The effect of temperature on the molecular arrangement within the monolayer near maximum uptake is shown in Figs.\ref{fig:hbonds-l8}(a-d), where the Hydrogen-bond network at various temperatures is depicted. A highly ordered H-bond network is observed, promoted by the narrow confinement, which forces the molecular plane of the water molecules to align parallel to the adsorbent layer. As expected, increasing temperature disrupts or "melts" this network $-$ this is specially visible at $400$~K, and is the reason behind the observed reduction in the maximum uptake observed in Fig.~\ref{fig:hyst-l8}(d). The hydrogen bonds were identified following the method described in Ref.~\cite{Sudheer2017}, with the cutoff radius and angle defined as $r_c = 3~\AA$ and $\theta_c = \pi$, respectively. Furthermore, the general configuration of these networks aligns with findings in the literature \cite{Yang2020,GaoLi,Leoni2021, Kapil2022}, although our system is laterally infinite, and we worked with relatively lower pressures.

\begin{figure}[h!]
   \centering
 \includegraphics[width=0.45\textwidth]{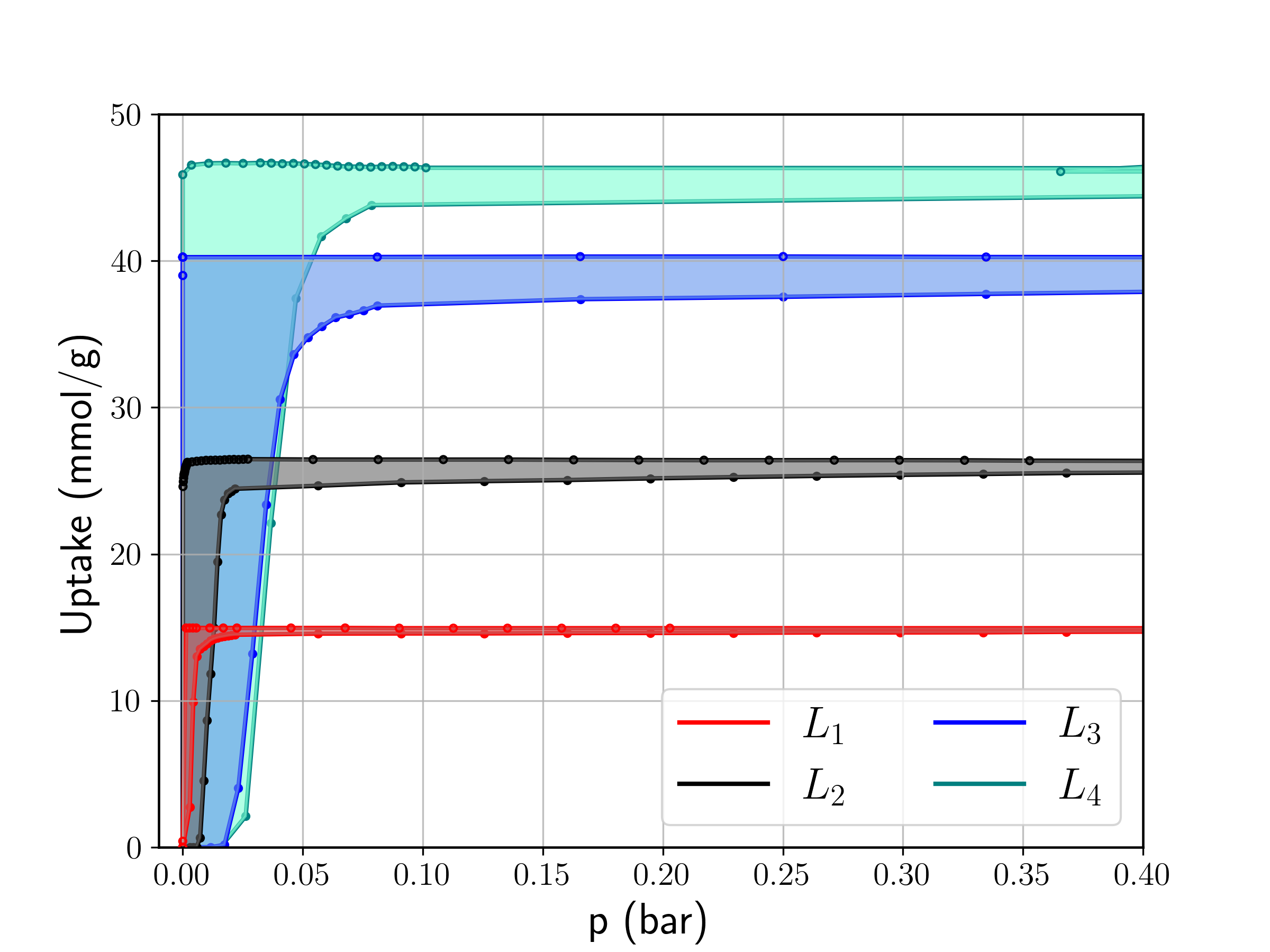}
    \caption{Isotherms at $T = 300$~K and pressure change rate $\kappa_{\rm f} = 4.83 \times 10^{-9}$ bar/MCs for the forward path uptake for all graphene-graphene distances considered.}
    \label{fig:layers}
\end{figure}

Next, we extend our analysis to the results obtained from the bilayer system ($L_2 = 10~\AA$). As expected, the maximum uptake per gram of adsorbent is nearly doubled compared to the monolayer (Fig.~\ref{fig:hyst-l8}(e-h)). However, at any given temperature, the H-bond network in the bilayer near maximum uptake (Fig.~\ref{fig:hbonds-l8}(e-h)) is more disordered than in its monolayer counterpart (Fig.~\ref{fig:hbonds-l8}(a-d)). We attribute this effect to the added rotational freedom of the water molecules in the bilayer, which makes their planes less constrained to align parallel to the graphene sheets. This increasing disorder in the H-bond network is expected to intensify with additional water layers, which also causes the maximum uptake per gram of adsorbent to grow at a sublinear rate with the number of water layers, as shown in Fig.~\ref{fig:layers}.

%(**se conseguir tomar a média de S=(1/2)[3*cos(theta)^2-1] do conjunto de moléculas, onde theta é o angulo entre a normal ao plano de cada molecula e a direção z teremos uma medida de ordenamento orientacional na direção z, S=1 todas as moleculas alinhadas com os grafenos, S=0 tudo abagunçado. No caso das multilayers e olhando a Fig 7(b-e), seria legal ter uma medida de S por camada de água **)

\begin{figure}[!h]
    \centering
    \includegraphics[width=0.45\textwidth]{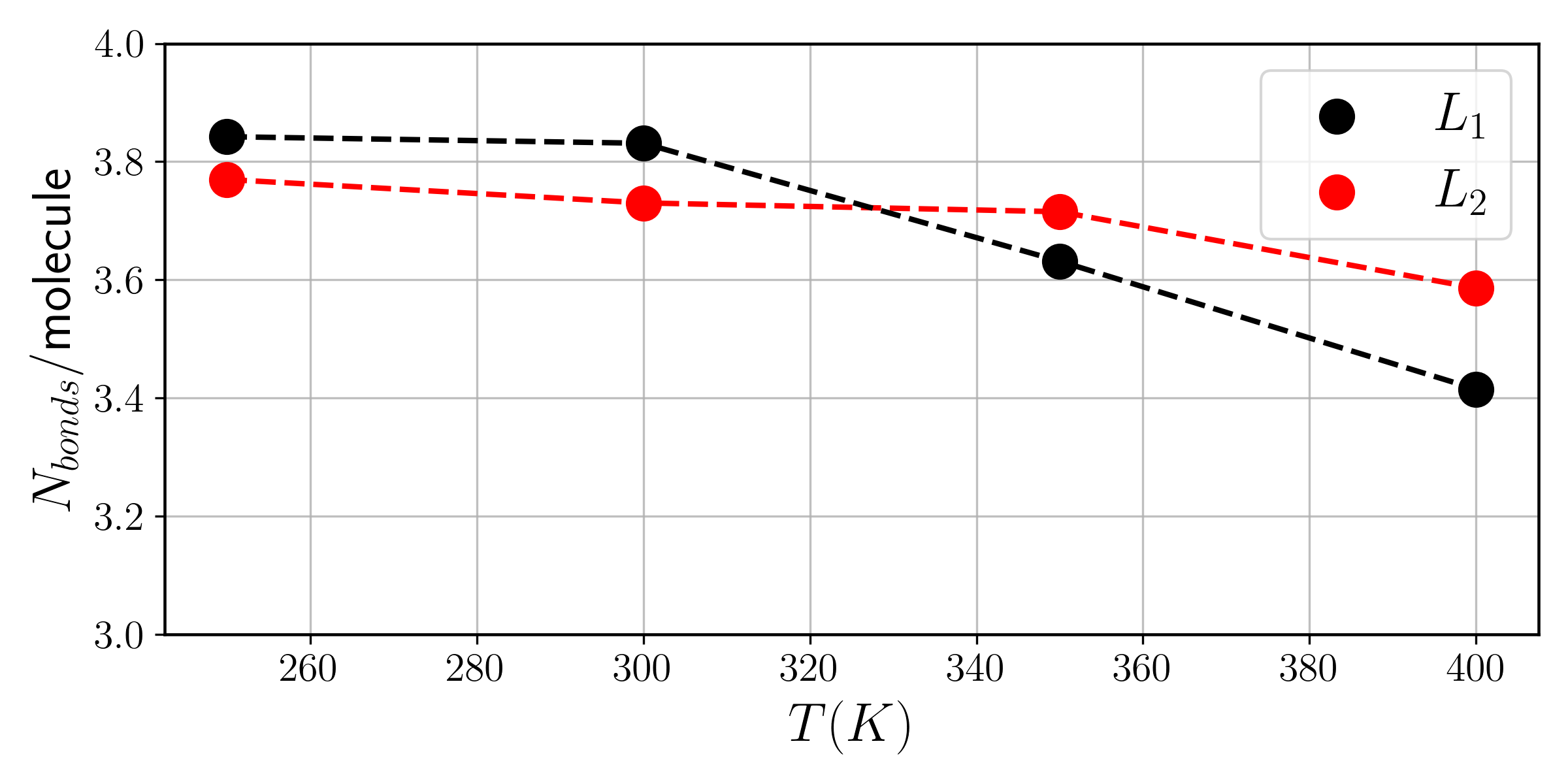}
    \caption{Average number of H-bonds per water molecule for $L_1$ (black) and $L_2$ (red) versus all temperatures investigated.}
    \label{fig:avbonds}
\end{figure}

In Fig.~\ref{fig:avbonds}, we compare the temperature effect on the average number of effective H-bonds per water molecule in the monolayer and bilayer cases. The results show a more pronounced temperature effect in disrupting the network in the monolayer, suggesting a more two-dimensional-like network in that case (therefore more susceptible to thermal effects), as compared with the bilayer.

\subsection{Temperature effect on the spatial and orientational arrangement of the water molecules}

\begin{figure}[h!]
\centering
    \includegraphics[width=0.52\textwidth]{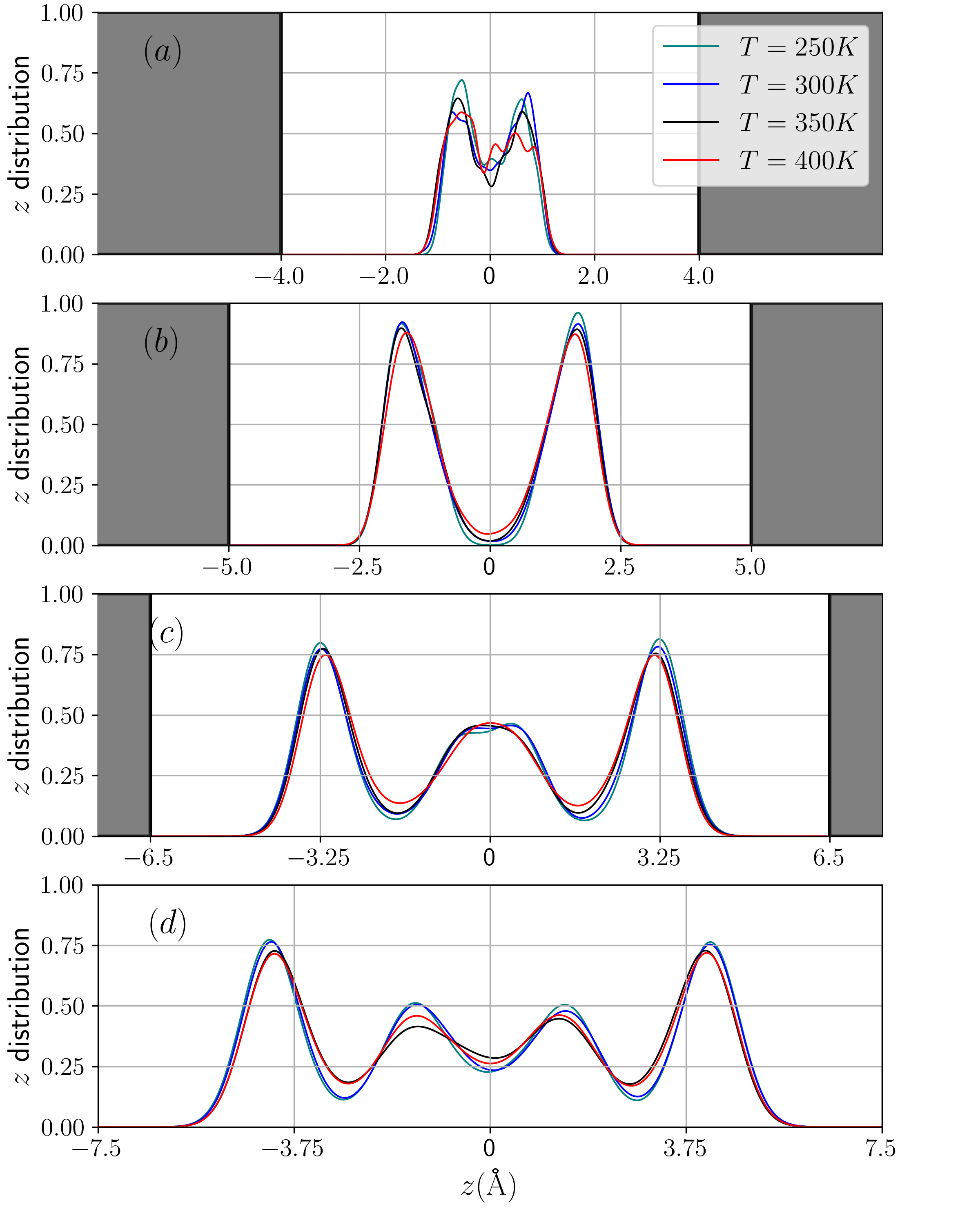}
    \caption{Distribution of water molecules across the distance separating the graphene sheets at four different temperatures for: (a) monolayer, (b) bilayer, (c) trilayer, and (d) quadrilayer systems. The distributions are normalized by the number of water layers, and the horizontal axis depicts the full inter-sheet distance.}
    \label{fig:hists}
\end{figure}

Figure~\ref{fig:hists} shows the water distribution along the direction perpendicular to the graphene sheets for all systems (mono-, bi-, tri-, and quadrilayers) at four different temperatures. The horizontal axis in each plot spans the entire inter-sheet region, and all axes are equally scaled. A noticeable feature is the non-uniformity in the distance between the graphene sheets and the nearest water layer. For instance, the monolayer peak is significantly farther from the graphene sheets compared to the other cases, which explains why the monolayer peak is beginning to split into two distinct layers. 
%and the non-uniformity in the width of the density peaks (e.g., ...)
This stems from our \emph{ad hoc} selection of the spacing between the graphene sheets. Moreover, in the tri and quadrilayer cases, there is a marked concentration of the water molecules in the surface layers, as opposed to the middle layers.

The primary effect of temperature is a slight broadening of the peaks, most notably observed as a small increase in density at the minima. This broadening occurs predominantly inward, except in the case of the mono-layer, where it extends both inward and outward.  An alternative view of the effect can be found in the supplementary material (Section 2, Figure S.2).
%(Section \ref{sdists}, Figure S.\ref{fig:dists_xz}).

To characterize the orientational arrangement of the water molecules within the graphene slits, we calculated the scalar order parameter for each water layer, defined by $S = (3/2) \langle \cos^2(\theta) \rangle - 1/2$, where $\theta_i$ is the angle between the director vector of the $i$-th water molecule and the axis perpendicular to the graphene sheets, and the average is over all water molecules within a given layer. $S\sim1$ in a given layer indicates that the water molecules are aligned predominantly parallel to the graphene sheets, whereas $S\sim 0$ indicates that their orientations are largely random. The results for the four cases near maximum uptake as a function of temperature are presented in Table \ref{tab:scalar_parameter}. 

Comparing the mono and bilayer cases, it is evident that the monolayer is more orientationally ordered than the bilayer at all temperatures, with temperature exerting a relatively stronger effect on the monolayer. These findings align with the discussion on the Hydrogen bond network in the previous section. 
In the tri- and quadrilayer systems, the surface layers—those adjacent to the graphene—are consistently more organized than the middle layer(s) across all temperatures. These observations collectively underscore the organizational effect of graphene layers on water molecules, a phenomenon known to significantly influence the flow of water between graphene sheets \cite{C3CS60253B}.

%The resulting measurements, averaged over the zz-position within the slit and shown in Fig.~\ref{fig:hists}, reveal that increasing temperature decreases the parallel alignment of water molecules with the graphene sheets. This effect is also evident in Fig.~\ref{fig:hbonds-l8}, leading to the conclusion that the diminished formation of H-bond networks is closely tied to the spatial orientation of the adsorbate layer.

\begin{table}[h!]
    \centering
    \caption{Values of the scalar nematic order parameter, $S$, measured for each water layer at different temperatures. Greater values of $S$ indicate water molecules more parallel to the graphene sheets.}
    \label{tab:scalar_parameter}
    \begin{tabular}{c|cccc}
        \hline
        %&&$S$\\
        %\hline\hline
        $ L $ & $250K$ &  $300K$& $350K$ & $400K$\\
        %$L$ & $\epsilon_s ~({\rm meV })$ &  $\epsilon_s ~({\rm meV })$ &  $\epsilon_s ~({\rm meV })$ &  $\epsilon_s ~({\rm meV })$\\
        \hline\hline
        1 & $0.7248$ &  $0.6951$& $0.5513$ & $0.4804$\\ \hline
        2 & $0.3992$ & $0.3971$ & $0.3772$ & $0.2996$\\ \hline
        3 - middle & $0.3081$ & $0.3112$  & $0.2390$& $0.2272$\\
         3 - surface & $0.4796$ & $0.4554$  & $0.3792$& $0.3469$\\ \hline
        4 - middle & $0.3030$& $0.3179$& $0.2581$ & $0.1940$\\
        4 -surface & $0.5182$& $0.4823$& $0.4086$ & $0.3515$\\
      
        \hline
    \end{tabular}
\end{table}

%Furthermore, Fig.~\ref{fig:hists}, together with Figs.~\ref{fig:hbonds-l8} and \ref{fig:hyst-l8}, leads to an interesting conclusion. First, the increase in temperature across all layers reduces the number of water molecules near the graphene sheets by spreading the inner layers. The broadening of the peaks in the distance distributions indicates that more molecules shift toward the middle and top of the slit, as illustrated in S.\ref{sdists}. Second, the increase in layer thickness significantly impacts the orientational arrangement of the adsorbate layers. For instance, in the monolayer, the nematic parameter is, on average, close to 1 (decreasing with temperature), whereas it becomes negative in the middle layers of the quadrilayer.

\subsection{Effect of the pressure change rate on the hysteresis loop}\label{time}

%Given that there is no direct translation between Monte Carlo steps and real time, and considering that adsorption is fundamentally an experimental process, we include a brief discussion on the former, as it has direct implications for the validity and relevance of the results presented thus far.
%Given that there is no direct translation between Monte Carlo steps and real time, and considering that adsorption is fundamentally an experimental process, we include a brief discussion on Monte Carlo steps, as they have direct implications for the validity and relevance of the results presented thus far.

%As mentioned in Section~\ref{sec:intro}, each point in all isotherm curves shown here were obtained from averages coming from $\tau_{MC}=6\times 10^5$ steps, and each curve always begun with the empty graphene structure and ended at $p= 10^-10 {\rm bar}$, with the system being emptied again. 

We examined the filling/emptying hysteresis loop of the structures under different  pressure change rates.

The rates were controlled by varying the waiting time $\tau_{\rm MC}$, measured by the number of MC steps at each new pressure value, and the constant pressure increment/decrement ($\Delta p$). We considered $\kappa = \Delta p/\tau_{\rm MC} = 7.65\times 10^{-7}~{\rm bar/MCs}, 1.5\times 10^{-7}~{\rm bar/MCs}, 1.15\times 10^{-7}~{\rm bar/MCs}$.

%As mentioned in Section~\ref{sec:intro}, each point in the isotherm curves presented here was obtained from averages over $\tau_{\rm MC} = 6 \times 10^5$ Monte Carlo steps. Each curve always began with the empty graphene structure and ended at $p = 10^{-10}~{\rm bar}$, after which the system was emptied again. This number of steps was chosen based on a comparison between curves of the same isotherm (with equal pressure range, temperature, and layer size) measured with different steps per pressure, as shown in Fig.~\ref{fig:times}.
\begin{figure}[!h]
    \centering
    \includegraphics[width=0.4\textwidth]{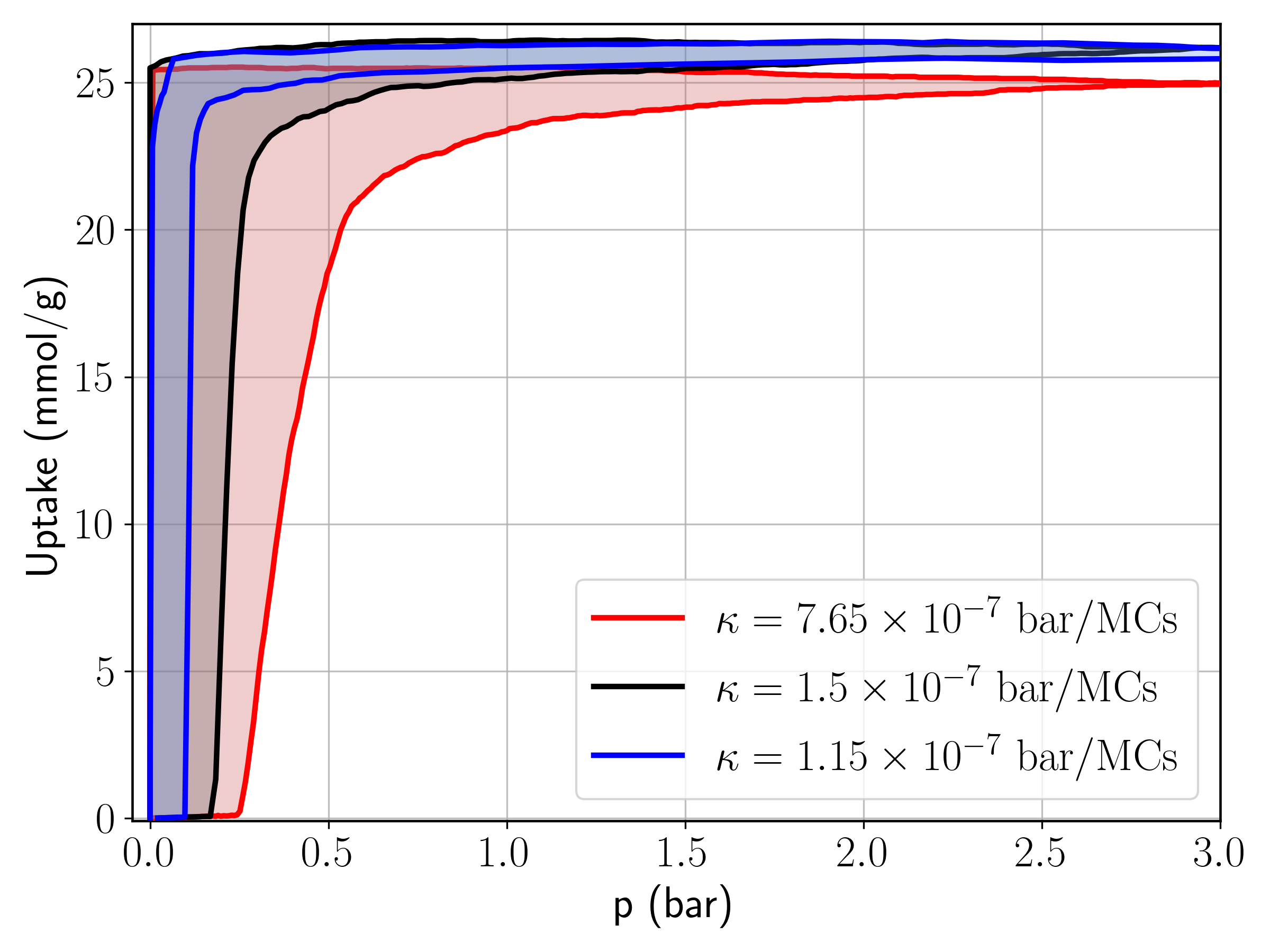}
    \caption{Forward and backward isotherms  at $T = 350$~K for the bilayer structure and for three different pressure change rates. The shaded region is bounded by the two isotherms.} 
    \label{fig:times}
\end{figure}

 The results, shown in Fig.~\ref{fig:times} for a bilayer at 350~K, display a marked characteristic of a first-order phase transition. Specifically, the slower the rate of pressure change, the smaller the hysteresis loop width. In the limit of infinitely slow pressure variation, we would observe a discontinuous filling/emptying jump at the coexistence pressure corresponding to that temperature. 

In all cases, irrespective of the selected rate, we observed the nucleation of water 'seedlings,' which served as initiation points for the phase transition via nucleation, as shown in S.1. This, together with the results from Fig.~\ref{fig:times}, strongly suggests that the adsorption of water in infinite graphene slits is better described as a first-order phase transition.

\subsection{Mean-field results}\label{sec:mf}

The mean-field treatment of the lattice model shown in Fig.~\ref{fig:modelo}, fully explained in \ref{ap2}, assumes equilibrium with an external reservoir at temperature $T$ and chemical potential $\mu$. We associate a variational occupancy $x_\alpha$ to all sites in the $\alpha$-th layer, thus exploiting the lateral symmetry. The top-bottom symmetry further restricts the number of independent variational parameters. The equation for the extremum of the mean-field free energy provides the $\mu$ and $T$ dependence of these occupations, from which adsorption isotherms can be obtained. In the case of multiple solutions, the mean-field free energy must be consulted to determine stability.

In the mono- and bi-layer cases, a single variational occupancy $x_s$ suffices. The mean-field equation of state is given by \cite{Donogue_1998}:
\begin{equation}\label{MonoBi}
\mu = kT \log \left( \frac{x_s}{1-x_s}\right) + \left\{ \begin{array}{c} 4 \\ 5 \end{array} \right\} \epsilon x_s + \left\{ \begin{array}{c} 2 \\ 1 \end{array} \right\} \epsilon_s,
\end{equation}
with the upper (lower) values for the mono- (bi-) layer case, and $\epsilon~(\epsilon_s)$ being the interaction of the adsorbate molecule with a nearest neighbor (the adsorbent surface).

In the tri- and quadri-layer cases, two variational occupancies are needed: $x_s$ for the surface layer sites and $x_m$ for the middle layer sites. The mean-field coupled system of equations of state is \cite{Donogue_1998}:
\begin{equation}\label{TriQuadri}
\begin{split}
\mu =&\, kT \log \left( \frac{x_s}{1-x_s}\right) + 4\epsilon x_s + \epsilon x_m + \epsilon_s, \\
\mu =&\, kT \log \left( \frac{x_m}{1-x_m}\right) + \left\{ \begin{array}{c} 4 \\ 5 \end{array} \right\} \epsilon x_m + \left\{ \begin{array}{c} 2 \\ 1 \end{array} \right\} \epsilon x_s,
\end{split}
\end{equation}
with the upper (lower) values for the tri- (quadri-) layer case.

The pattern is clear, and one can easily write the system of equations for a system with $L$ layers. It is also evident that the equilibrium occupations depend only on three parameters: $\mu/kT$, $\epsilon/kT$, and $\epsilon_s/kT$.

For future reference, we present the equation for the bulk (simple cubic lattice) case, which follows the same pattern:
\begin{equation}
\mu = kT \log \left( \frac{x}{1-x}\right) + 6\epsilon x.
\end{equation}

\begin{figure*}[h]
    \centering
  %  \begin{subfigure}[t]{0.32\textwidth}
   %     \centering
    %    \includegraphics[width=\textwidth]{Theta1.png}\label{fig:Theta1}
        % \caption{}
    %\end{subfigure}
    %\hfill
    %\begin{subfigure}[t]{0.32\textwidth}
     %   \centering
      %  \includegraphics[width=\textwidth]{Theta2.png}\label{fig:Theta2}
        % \caption{}
 %   \end{subfigure}
  %  \hfill
   % \begin{subfigure}[t]{0.32\textwidth}
    %    \centering
     %   \includegraphics[width=\textwidth]{Theta3.png}
        % \caption{Trilayer case.}
      %  \label{fig:Theta3}
   % \end{subfigure}
   \includegraphics[width=\textwidth]{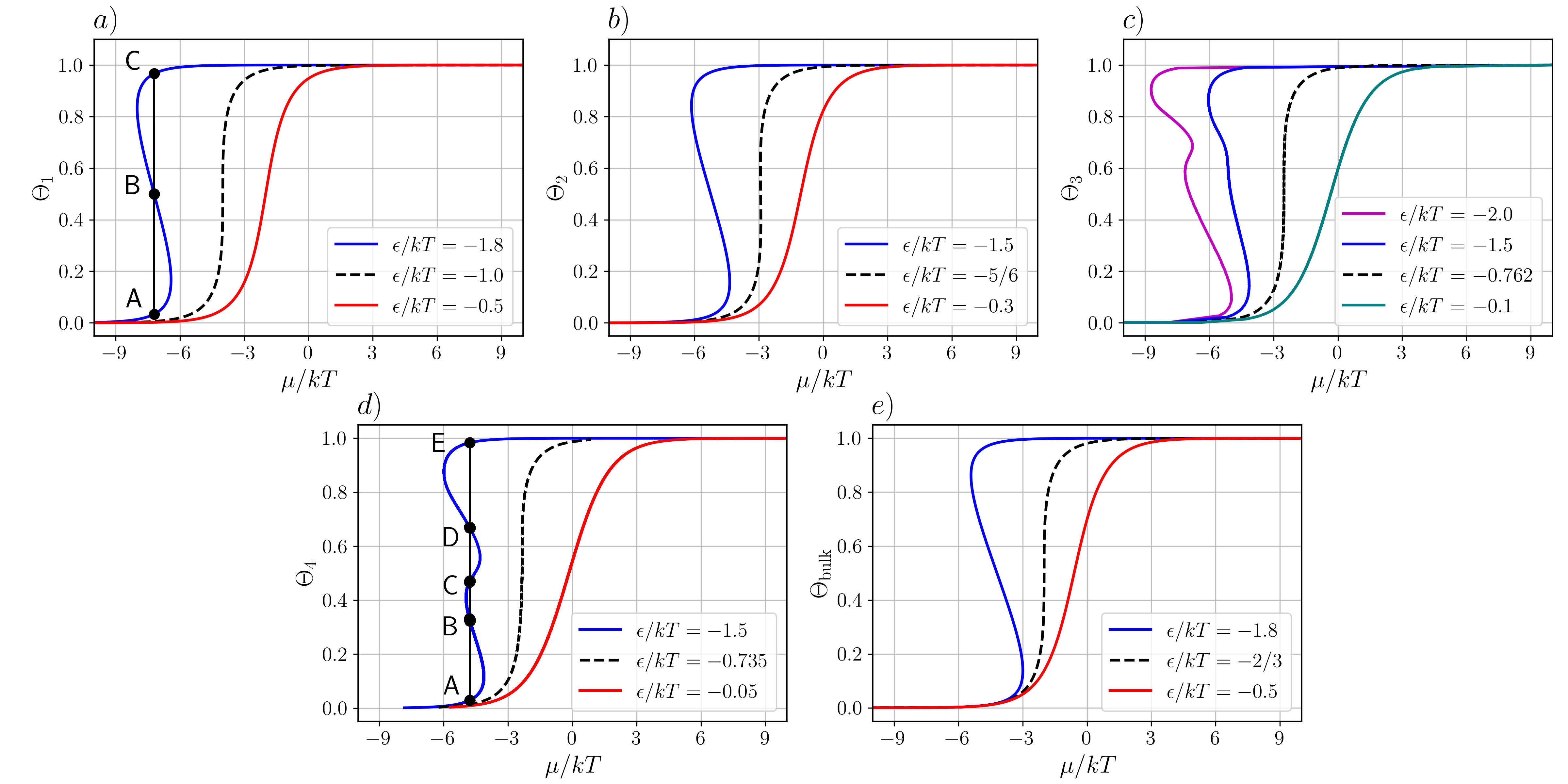}
    \caption{In all cases, $\epsilon=\epsilon_s$. $\Theta$ is the average site occupation, and the dashed curve corresponds to the critical isotherm. (a) Monolayer case: The mean-field free energy along the vertical line is shown in Fig. \ref{fig:Omega14}$(a)$. From that figure, we identify A as stable, B as unstable, and C as metastable; $(b)$ Bilayer case; $(c)$ Trilayer case; $(d)$ Quadrilayer case: the mean field free energy, as a function of $x_s$ and $x_m$, and for the chemical potential corresponding to the vertical line, is shown in Fig.~\ref{fig:Omega14}(b). From that figure, we identify A as metastable (local minimum), B, C, and D as unstable (saddle
points/local maximum), and E as stable (global minimum); $(e)$ Bulk case.}
    \label{fig:Thetas}
\end{figure*}

%\begin{figure}[h]
 %   \centering
  %  \begin{subfigure}[t]{0.32\textwidth}
   %     \centering
    %    \includegraphics[width=\textwidth]{Theta4.png}\label{fig:Theta4}
        % \caption{}
  %  \end{subfigure}
   % \hfill
    %\begin{subfigure}[t]{0.32\textwidth}
     %   \centering
      %  \includegraphics[width=\textwidth]{ThetaBulk.png}\label{fig:ThetaBulk}
        % \caption{}
    %\end{subfigure}
  %  \includegraphics[width=0.5\textwidth]{figure_9.png}
  %  \caption{In both cases, $\Theta$ is the average site occupation, and the dashed curve corresponds to the critical isotherm. (a) Quadrilayer case ($\epsilon=\epsilon_s$): The mean-field free energy along the vertical line is shown in Fig. \ref{fig:Omega14}(b). From that figure, we identify A as stable (global minimum), B, C, and D as unstable (saddle points), and E as metastable (local minimum); (b) Bulk case. }
   % \label{fig:Theta34bulk}
%\end{figure}

%In Figs.~\ref{fig:Theta123} and \ref{fig:Theta34bulk}, we present the average site occupation for the mono- to quadri-layer cases, e.g. $\Theta_2 $  (2xs+xm)/3, as well as for the bulk case, ...

\begin{figure*}[!h]
    \centering
   % \begin{subfigure}[t]{0.32\textwidth}
    %    \centering
     %   \includegraphics[width=\textwidth]{Omega1.png}\label{fig:Omega1}
        % \caption{}
    %\end{subfigure}
    %\hfill
    %\begin{subfigure}[t]{0.32\textwidth}
     %   \centering
      %  \includegraphics[width=\textwidth]{Omega4.png}\label{fig:Omega4}
        % \caption{}
    %\end{subfigure}
 % \includegraphics[width=0.4\textwidth]{figure_10.png}
   \includegraphics[width=0.4\textwidth]{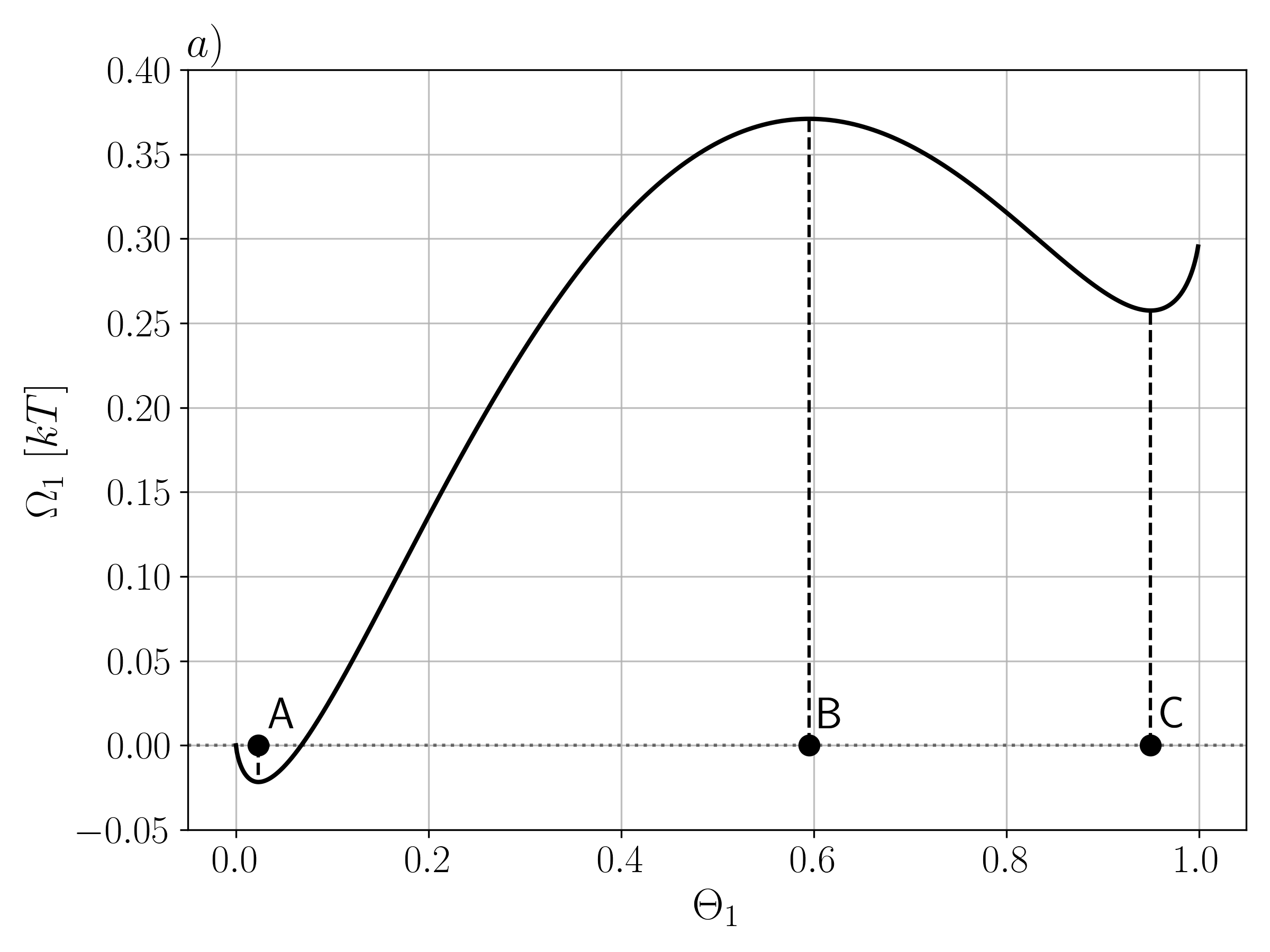}
    \includegraphics[width=0.4\textwidth]{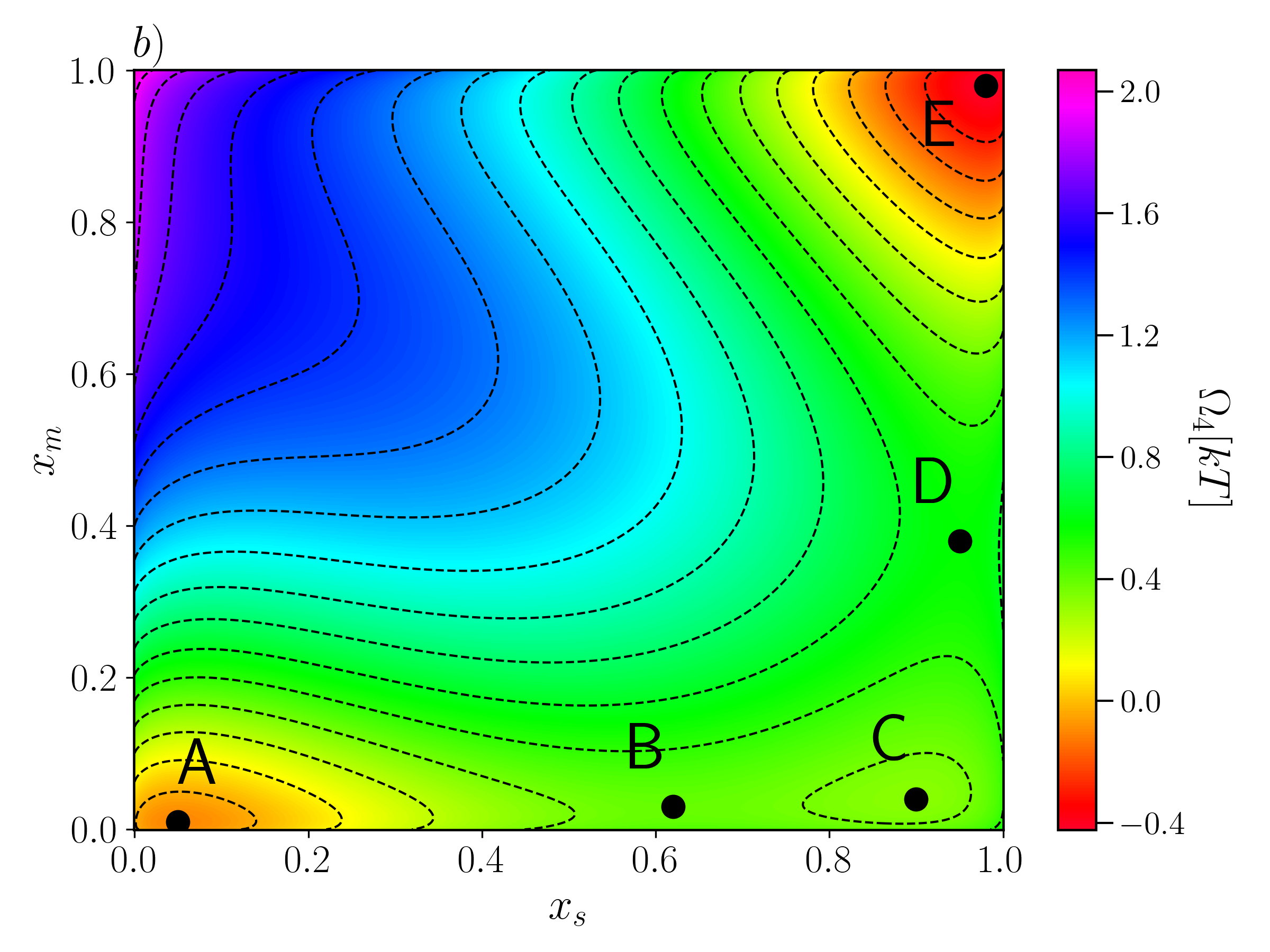}
    \caption{(a) The monolayer mean-field free energy per surface area for $\epsilon/kT=\epsilon_s/kT=-1.8$ and $\mu/kT=-7.5$. The labeled points are the extrema, solutions to Eq. \eqref{MonoBi}. A is stable, B is unstable, and C is metastable. These points are identified in the corresponding isotherm of Fig. \ref{fig:Thetas}(a). (b) The quadrilayer mean-field free energy per surface area for $\epsilon/kT=\epsilon_s/kT=-1.5$ and $\mu/kT=-4.7$. $x_s$ ($x_m$) is the average site occupation at the surface (middle) layers. The labeled points are the extrema, solutions to Eq. \eqref{TriQuadri}. A is a local minimum (metastable), B, C and D are saddle-points/local maximum (unstable) and E is the global minimum (stable). These points are identified in the corresponding isotherm of Fig. \ref{fig:Thetas}(d), where $\Theta_4=(2x_s+2x_m)/4$. }
    \label{fig:Omega14}
\end{figure*}

Fig. \ref{fig:Thetas} shows the average site occupation versus $\mu/kT \propto \log(p)$, for the mono to quadrilayer, $\Theta_L$, e.g., $\Theta_3/3=(2x_s+x_m)/3$, as well as the bulk case, in the particular case where the surface attraction equals the intermolecular attraction, $\epsilon=\epsilon_s$. The black dashed lines correspond to the critical isotherm. Below the critical temperature, Eqs. \eqref{MonoBi} and \eqref{TriQuadri} admit multiple solutions for a given value of $\mu$, meaning that the mean-field free energy has multiple extrema in the variational parameter space. This is illustrated in Figs. \ref{fig:Omega14}(a) and \ref{fig:Omega14}(b) for the monolayer and quadrilayer cases.

The critical temperature and chemical potential can be found analytically in the bulk case ($L=\infty$) and in the mono- and bi-layer cases, as shown in Table \ref{tab:analytical}. We find that: (i) as expected, the phase transition only occurs for attractive intermolecular interactions; (ii) surface interaction does not affect the critical temperature, which is proportional to $|\epsilon|$; (iii)  confinement lowers $T_c$ with respect to the bulk value; (iv) the critical chemical potential depends linearly on both $|\epsilon|$ and $\epsilon_s$, with surface attraction lowering $\mu_c$; (v) the critical pressure, estimated as $p_c \sim e^{\mu_c/kT_c}$, is exponentially reduced by surface attraction compared to its bulk value, with the effect further enhanced by confinement.

\begin{table}[h!]
    \centering
    \caption{Analytical expressions for the mean-field critical temperature and chemical potential ($\epsilon<0$)}\label{tab:analytical}
    \begin{tabular}{c|ccc}
        \hline \hline
        $L$ & $kT_c$ & $\mu_c$ & $\log p_c \sim \mu_c/kT_c$ \\
        \hline
        1 & $|\epsilon|$ & $-2\,|\epsilon|+2\,\epsilon_s$ & $-2+2\,{\epsilon_s}/{|\epsilon|}$ \\
        2 & $1.2\,|\epsilon|$ & $-2.4\,|\epsilon|+1.92\,\epsilon_s$ & $-2+1.6\,{\epsilon_s}/{|\epsilon|}$ \\
        $\infty$ & $1.5\,|\epsilon|$ & $-3\,|\epsilon|$ & $-2$ \\
        \hline \hline
    \end{tabular}
\end{table}

The tri- and quadri-layer cases do not allow for an analytical solution. We illustrate the numerical results for the specific case of $\epsilon_s=\epsilon<0$ in Table \ref{tab:numerical}. The trends discussed above, particularly the lowering of $T_c$ and the exponential lowering of $p_c$ with confinement, are all evident.
\begin{table}[h!]
    \centering
    \caption{Mean-field critical temperature and chemical potential for different numbers of layers in the particular case of $\epsilon=\epsilon_s<0$}
    \label{tab:numerical}
    \begin{tabular}{c|lll}
        \hline\hline
        $L$ & $kT_c/|\epsilon|$ & $\mu_c/|\epsilon|$ & $\log p_c \sim \mu_c/kT_c$ \\
        \hline
        1 & $1$ & $-4$ & $-4$ \\
        2 & $1.2$ & $-4.32$ & $-3.6$ \\
        3 & $1.312...$ & $-3.29...$ & $-2.50...$ \\
        4 & $1.361...$ & $-3.19...$ & $-2.34...$ \\
        $\infty$ & $1.5$ & $-3$ & $-2$ \\
        \hline\hline
    \end{tabular}
\end{table}

%BDDT \cite{BDDT_1940}
%Interaction Energy \cite{Donogue_1998, Hamada_2012}
%Ono-Kondo \cite{OnoKondo}

\subsection{Mean-field parameters extracted from Molecular Dynamics isotherms}

% \begin{figure}[!h]
%   %  \centering
% \flushleft
%    \includegraphics[width=4.3cm]{fitzero.png} 
%    \includegraphics[width=4.3cm]{fitright.png} 
%    % \caption{2 Figures side by side}%
%     %\label{fig:example}%
%     \caption{Fitting results of MD data points (scatter plots) and resulting analytical function (red dashed curves) for $a)$ $\Theta_2\to 0$ and $b)$ $\Theta_2 \to 1$ in the $L_2$ system and $T=350K$. The values obtained for $\epsilon_s$ and $\epsilon$ can be found in Tables \ref{tab:fittings_es} and \ref{tab:fittings_e}, respectively.}
%     \label{fig:mfcomp}
% \end{figure}

The fitting of the MD data was performed using Eqs.\ref{eq:lowp} for the values of $\epsilon_s$, and \ref{eq:highp} for the values of $\epsilon$. An illustration of typical fitting result is shown in the supplementary material, in Section 3, Fig.~3, and the results obtained are summarized in Tables \ref{tab:fittings_es} and \ref{tab:fittings_e}.

The absolute values of both energies is seen to increase with temperature, a result of the limitations of a rigid lattice model to capture the physics of a system where the molecular orientations and their proximity to each other and to the adsorbate are all affected by the temperature as see in Table \ref{tab:scalar_parameter} and in Fig. \ref{fig:hists}.

 Nonetheless, the value of $\epsilon_s$
  obtained for the monolayer is roughly consistent with those reported in the literature \cite{Hamada_2012}, while the water-water interaction energy $\epsilon$ is somewhat lower than the typically quoted value for the hydrogen bond in water ($\sim 200$ meV). Since the lattice model attempts to encapsulate in a single energy parameter all energies associated with different relative orientations between water molecules and graphene, in the case of $\epsilon_s$, as well as between water molecules themselves, in the case of $\epsilon$, this may indicate that the former interaction is less sensitive to precise relative orientations compared to the latter, which makes physical sense.

  The significantly lower absolute values of $\epsilon_s$ in the monolayer case as compared to the other cases are easily explained by the greater distance of the water molecules to the graphene sheets in that case compared to the others, as seen in Fig. \ref{fig:hists}.

\begin{table}[h!]
    \centering
    \caption{Values of the model parameter $\epsilon_S$ obtained by fitting molecular dynamics isotherms in the low-pressure limit.}
    \label{tab:fittings_es}
    \begin{tabular}{c|llll}
        \hline
        &&$-\epsilon_s ~({\rm meV })$\\
        \hline\hline
        $ L $ & $250K$ &  $300K$& $350K$ & $400K$\\
        %$L$ & $\epsilon_s ~({\rm meV })$ &  $\epsilon_s ~({\rm meV })$ &  $\epsilon_s ~({\rm meV })$ &  $\epsilon_s ~({\rm meV })$\\
        \hline
        1 & $167.07$ & $171.95$ & $185.35$ & $201.98$ \\
        2 & $264.19$ & $279.09$ & $297.93$ & $333.58$\\
        3 & $249.68$ & $274.86$  & $301.71$& $335.07$\\
        4 & $242.01$& $260.97$& $289.89$ & $323.02$\\
      
        \hline\hline
    \end{tabular}
\end{table}

\begin{table}[h!]
    \centering
    \caption{Values of the model parameter $\epsilon$ obtained by fitting molecular dynamics isotherms in the high-pressure limit.}
    \label{tab:fittings_e}
    \begin{tabular}{c|llll}
        \hline
        &&$-\epsilon ~({\rm meV })$\\
        \hline\hline
        $ L $ & $250K$ &  $300K$& $350K$ & $400K$\\
        %$L$ & $\epsilon_s ~({\rm meV })$ &  $\epsilon_s ~({\rm meV })$ &  $\epsilon_s ~({\rm meV })$ &  $\epsilon_s ~({\rm meV })$\\
        \hline
        1 & $29.31$ & $48.43$ & $62.21$ & $62.75$ \\
        2 & $50.61$ & $63.83$ & $65.84$ & $67.76$\\
        3 & $70.51$ & $81.83$  & $88.53$& $91.01$\\
        4 & $75.12$& $80.91$& $95.24$ & $99.91$\\
      
        \hline\hline
    \end{tabular}
\end{table}

% We observe that the fitting results are consistent with several values reported in the literature \cite{hbondsok, Hamada_2012, LIAO2022154477}. However, a distinct contrast between the monolayer and the other systems is apparent and has been discussed throughout this paper. Moreover, we attribute the temperature-dependent variations in surface energy—and consequently, the adsorbate-adsorbate interaction energy—to the increased confinement of the adsorbate layer. This confinement leads to an expansion of the layer, thereby reducing the distance between the adsorbate and the graphene surface, as demonstrated in Section 2 of the supplementary material. Additionally, we believe that the choice of interlayer distance plays a critical role in the fitting values obtained. This measure can be found in S.\ref{sdists}

\section{Conclusion}
\label{conclusion}

We conducted molecular dynamics (MD) simulations to investigate water adsorption on laterally unconfined graphene slits, utilizing the Grand Canonical Monte Carlo (GCMC) method implemented in LAMMPS. The isotherms were generated by incrementally increasing the reservoir pressure from zero to the maximum uptake, followed by a gradual decrease in pressure until the system was nearly devoid of water.

All isotherms produced displayed hysteresis, indicative of a genuine first-order phase transition, distinct from the surface tension effects observed in finite slits. Notably, the hysteresis loop widened with an increased rate of pressure change, a characteristic feature of first-order phase transitions.

Further analysis of the water molecular arrangement at maximum uptake revealed the formation of hydrogen bond networks in both monolayer and bilayer systems. We characterized the orientation of water molecules using the scalar nematic order parameter, observing that graphene sheets induce a parallel orientation of water molecules in the nearest layers. This alignment is a significant contributor to the formation of the hydrogen bond network. Conversely, the orientation-driven effect diminishes in the deeper layers of water.

This orientational ordering induced by nano-confinement not only enhances adsorption uptake but is also expected to facilitate increased water flow during transport processes.

In addition, we examined the Ono-Kondo lattice model of water confined between two infinite parallel sheets at the mean-field level. Our analytical and numerical results indicate a reduction in both the critical temperature and pressure compared to bulk values, with a particularly notable decrease in critical pressure. This reduction was quantified by the factor $\exp^{-\alpha|\epsilon_s|/|\epsilon|}$, where $\epsilon_s$ and $\epsilon$ denote the adsorbate-adsorbent and inter-adsorbate interactions, respectively, and $\alpha$
varies from 0 in the bulk limit to 2 in the monolayer case, see Tables \ref{tab:analytical} and \ref{tab:numerical}. 

In summary, our study elucidates the phase transition mechanisms underlying water adsorption in extended nanopores, findings that are likely applicable to other adsorbates and adsorbents within similar geometries.

%The effect of temperature in the formation of hydrogen bonds is well established in literature \cite{Nelson2005,Marchal2006TheHB}, and can be seem in across all four panels in Fig.~\ref{fig:hbonds-l8}, as increasing temperature (at the same pressure and layer size) reduces the number of bonds per molecule (see Fig.~\ref{fig:hbonds-l8}.\textit{e)}), a result also expected for the rigid water model utilized in this work~\cite{Ohtaki2003,Mizan1996, Swiatla-Wojcik2008}. Additionally, our results for the monolayer further reinforce the idea that hydrogen bonding plays a key role in the process of both adsorption and desorption~\cite{GILORMINI2018164,Chepurnoi1984}, as the reduction in number of bonds also accompanies both a reduction in uptake as the slit is loaded, and a less "persistent" saturation configuration, as the less connected network of h-bonds further reduces the necessary energy necessary for a water molecule to escape the slit. Can be understood as a direct consequence of the combination of the effects confinement and the effects introduced by the interactions due to the water molecules.  

\section*{Acknowledgements}

The authors would like to thank CNPq and CAPES for financial support and CCJDR for granting access to the HPC Cluster Coaraci, made available under FAPESP grants 2013/08293-7 and 2019/17874-0.


\appendix

%\section{Supplementary Information}
%\label{ap1}
\section{Derivation of Mean-field results}
\label{ap2}
\subsection{Mean-Field Treatment of the Lattice Gas Inside a Planar Slit}
Refer to Fig. \ref{fig:modelo}. The state of a given site can be $n_i=\{0,1\}$, and the full microstate is denoted
$\vn = (n_1, n_2, \ldots, n_{LM})$, with $M$ square lattice sites (with lateral periodic boundary conditions) stacked in $L$ layers.

The energy of a given microstate is:
\begin{equation}
    E_\vn = \frac{\epsilon}{2} \sum_{\langle i,j \rangle} n_i n_j + \epsilon_s \sum_{i\in {\cal S}} n_i - \mu \sum_i n_i,
\end{equation}
where the first sum is over all nearest neighbors and the second sum is restricted to sites on the top and bottom layers.

The variational mean-field probability distribution to be used, denoted $\rho_\vn(\vx)$, assigns a variational occupancy $x_\alpha$ to all sites belonging to the $\alpha$-th layer, thereby exploring the lateral symmetry. The top-bottom symmetry further restricts the number of independent variational parameters to $\lceil L/2 \rceil$.

The mean-field (MF) free energy is given by \cite{chaikinlubensky}:
\begin{equation}
    \Omega(\vx) = \sum_\vn \rho_\vn(\vx) E_\vn + kT\sum_\vn \rho_\vn(\vx) \log \rho_\vn(\vx).
\end{equation}

For our model and in terms of our choice of variational parameters (we denote $x_s$ as the variational occupancy of the surface layers and use the index $\alpha$ to label the layers):
\begin{equation}
\begin{split}
    \frac{\Omega(\vx)}{M} =&\,
    \frac{\epsilon}{2} \left[ \sum_{\langle \alpha, \alpha' \rangle} x_\alpha x_{\alpha'} + 4\sum_\alpha x_\alpha^2 \right] + 2\epsilon_s x_s - \mu \sum_\alpha x_\alpha \\
    & + kT \sum_\alpha x_\alpha \log x_\alpha + (1-x_\alpha) \log (1-x_\alpha).
\end{split}
\end{equation}

For instance, in the case of a quadri-layer, we obtain (denoting $x_m$ as the variational occupancy of the two middle layers):
\begin{equation}
\begin{split}
    \frac{\Omega_4(x_s, x_m)}{M} =&\, 
    \epsilon \left[ 2x_s x_m + 4x_s^2 + 5x_m^2 \right] + 2\epsilon_s x_s - 2\mu (x_s + x_m) \\
    & + 2kT \left[ x_s \log x_s + (1-x_s) \log (1-x_s)\right.\\
    &\, \left. + x_m \log x_m + (1-x_m) \log (1-x_m) \right]
\end{split}
\end{equation}
The equations for the extrema of this function are shown in Eq.~\eqref{TriQuadri}, and a plot for a particular choice of $(\epsilon, \epsilon_s, \mu)$ is shown in Fig.~\ref{fig:Omega14}(b).

\subsubsection{The Isotherms at Low and High Pressures}

In the limit of low chemical potential, $\mu/kT \ll \epsilon_s/kT < 0$, the system is nearly empty, and we have (see Eq. \eqref{TriQuadri}):
\begin{equation}
\quad x_\alpha \sim  \begin{cases}
            e^{(\mu - \epsilon_s)/kT}, & \alpha = s\\
            e^{\mu/kT}, & \alpha \ne s
         \end{cases} 
\end{equation}

This implies\footnote{The monolayer case is special, see Eq.~\eqref{MonoBi}: $x_s = \Theta_1 \sim e^{\mu/kT}[e^{-2\epsilon_s/kT}]$.} 
\begin{equation}
\Theta_{L} \sim e^{\mu/kT}\, L^{-1} \left[ 2e^{-\epsilon_s/kT} + L - 2 \right].
\label{eq:lowp}
\end{equation}

The slope of the $\Theta$ vs. $e^{\mu/kT} \propto p$ curve at low pressures (Henry's constant) is strongly enhanced by surface attraction, but the effect is reduced with an increasing number of layers towards the bulk value ($L \to \infty$). The slope can be used to infer the value of the model parameter $\epsilon_s$ from an actual measurement.

Similarly, at large chemical potential, $\mu/kT \gg \max\{(5\epsilon + \epsilon_s)/kT, 6\epsilon/kT\}$, the system is nearly full, and we have (see Eq.~\eqref{TriQuadri}):
\begin{equation}
\quad x_\alpha \sim  \begin{cases}
            1 - e^{(5\epsilon + \epsilon_s - \mu)/kT}, & \alpha = s\\
            1 - e^{(6\epsilon - \mu)/kT}, & \alpha \ne s
         \end{cases} 
\end{equation}
This implies\footnote{The monolayer case is special, see Eq. \eqref{MonoBi}: $x_s = \Theta_1 \sim 1 - e^{(4\epsilon + 2\epsilon_s - \mu)/kT}$.} 
\begin{equation}
\Theta_{L} \sim 1 - e^{-\mu/kT}\, L^{-1} \left[ 2e^{(5\epsilon + \epsilon_s)/kT} + (L - 2)e^{6\epsilon/kT} \right].
\label{eq:highp}
\end{equation}
The behavior of $\Theta$ at large $e^{\mu/kT} \propto p$ can be used to infer 
the value of the model parameter $\epsilon$ from an actual measurement.

%\section{Appendix title 2}

\biboptions{sort&compress}
\bibliographystyle{elsarticle-num} 
\bibliography{references}

%\biboptions{sort&compress}
%\bibliographystyle{unsrt}


%% For example:

%% \bibitem[Aladro et al.(2015)]{Aladro15} Aladro, R., Martín, S., Riquelme, D., et al. 2015, \aas, 579, A101

%%\end{thebibliography}

\end{document}

\maketitle

\section{Evidence of Nucleation}
\label{snucl}
In this section, we show snapshots for the bilayer system, at $T = 300 K$ and pressure change rate $\kappa = 2.38\times 10^{-9}{\rm bar/MCs}$ in order to demonstrate the occurrence of nucleation during the adsorption process.

\begin{figure}[h!]
 %\hspace{-1cm}
 \centering
    \includegraphics[scale=0.7]{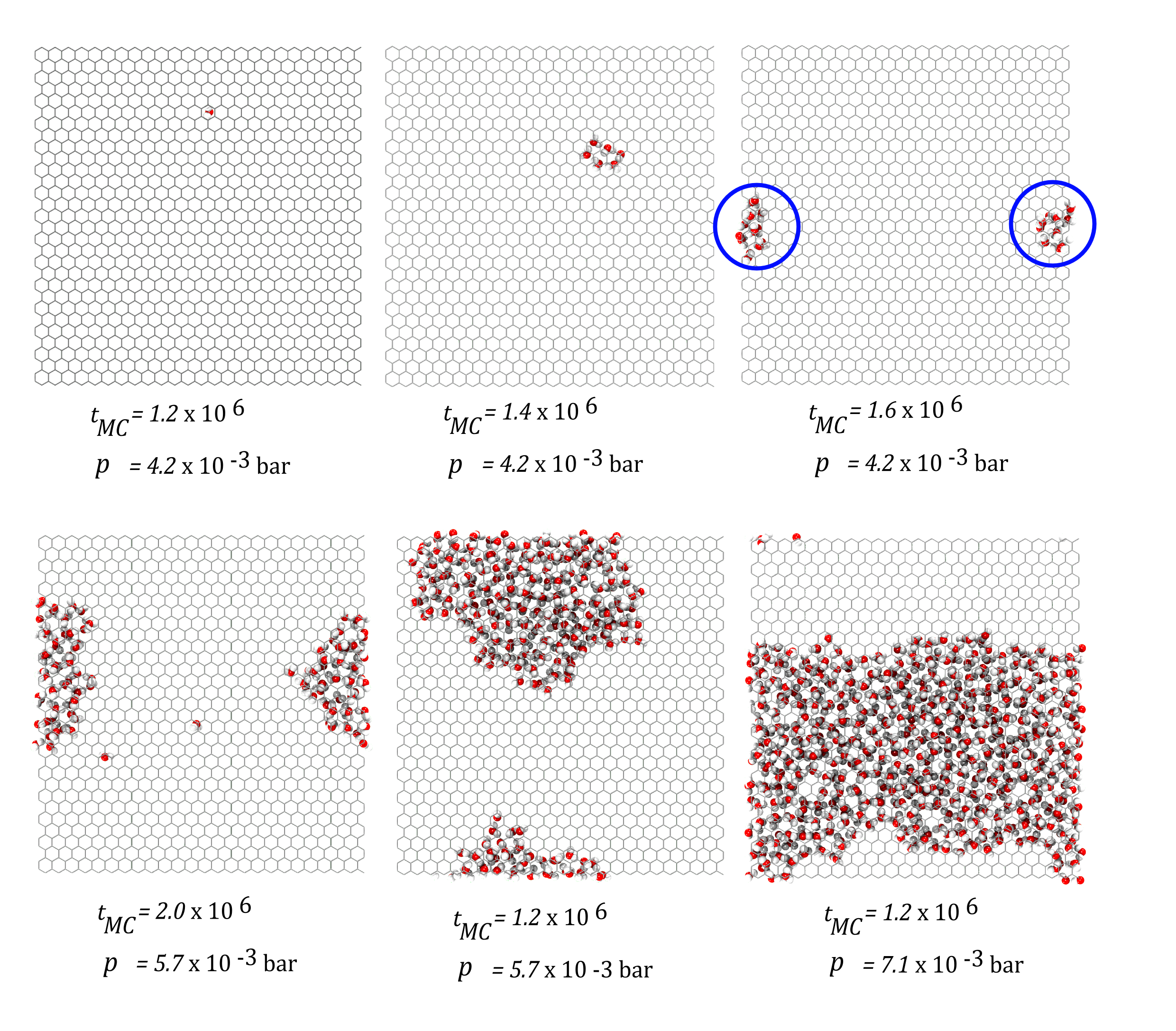}
    \caption{Snapshots of the bilyer, $T = 300K$ and $\kappa = 2.38 \times 10^{-9}$ bar/MCs. TThe blue circle marks the onset of the nucleation process, where a small cluster of water molecules forms, providing a nucleus to which subsequently adsorbed molecules will aggregate.}
    \label{nucl}
\end{figure}

\section{Water molecules distribution inside the graphene slit}
\label{sdists}

In this supplementary section, we illustrate the average spatial organization of the water molecules inside the graphene slit. In these illustrations, the water molecules (Oxygen and Hydrogen atoms) are not plotted in scale, and are colored in black, while the graphene sheets are colored in red.  

\begin{figure}[h!]
    \centering
    \includegraphics[width=1\linewidth]{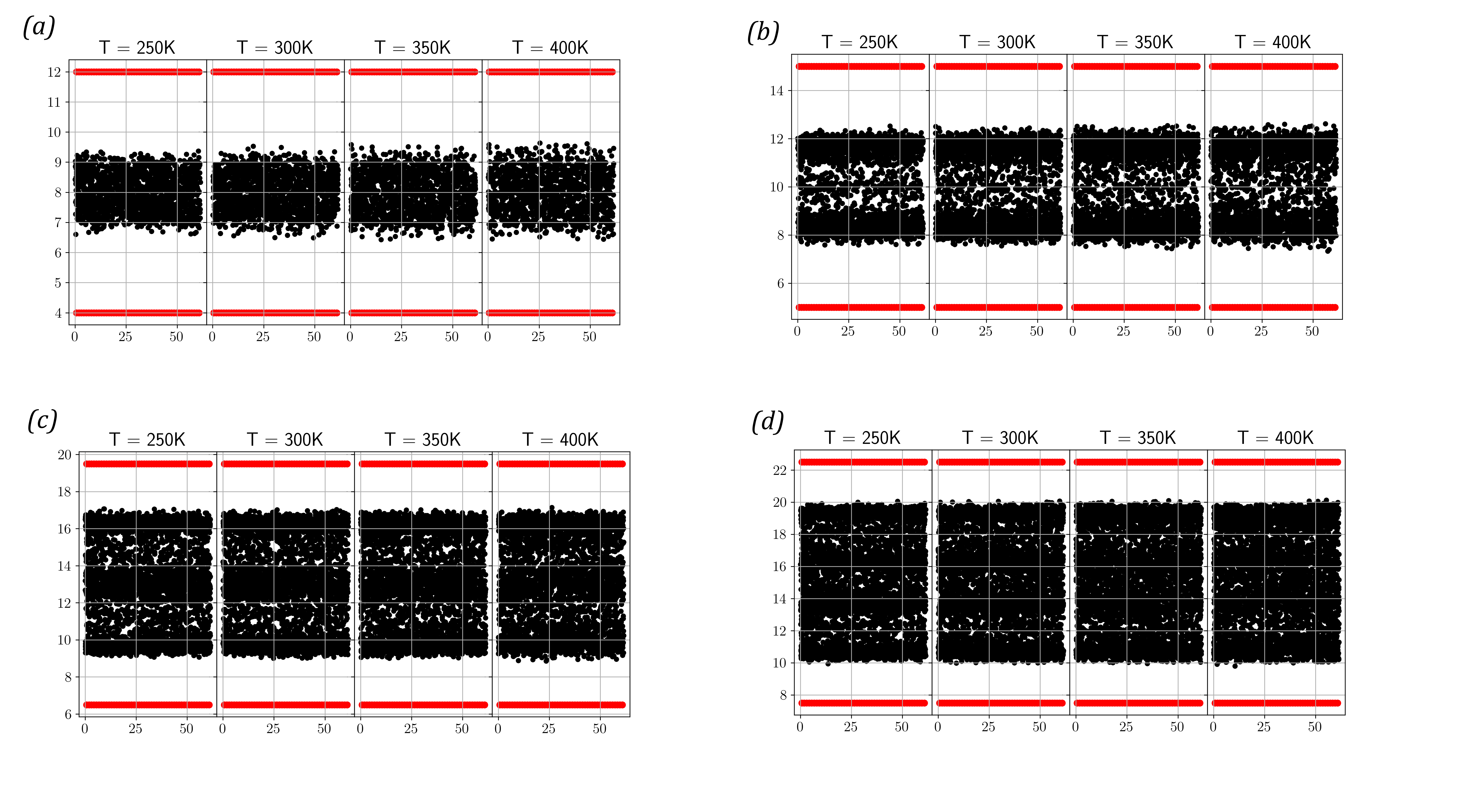}
    \caption{View of the $xz$ plane for all temperatures and the $(a)$ monolayer, $(b)$ bilayer, $(c)$ trilayer and $(d)$ quadrilayer. For easier visualization, Oxygen and Hydrogen atoms are scattered in black, while the Carbon atoms can be seen in red. The effect of the temperature on the average "amplitude" of the inner layers can be seen here. }
    \label{fig:dists_xz}
\end{figure}

\section{Typical mean-field fitting}
\label{smf}

 \begin{figure}[!h]
 \centering
% \flushleft
    \includegraphics[width=6cm]{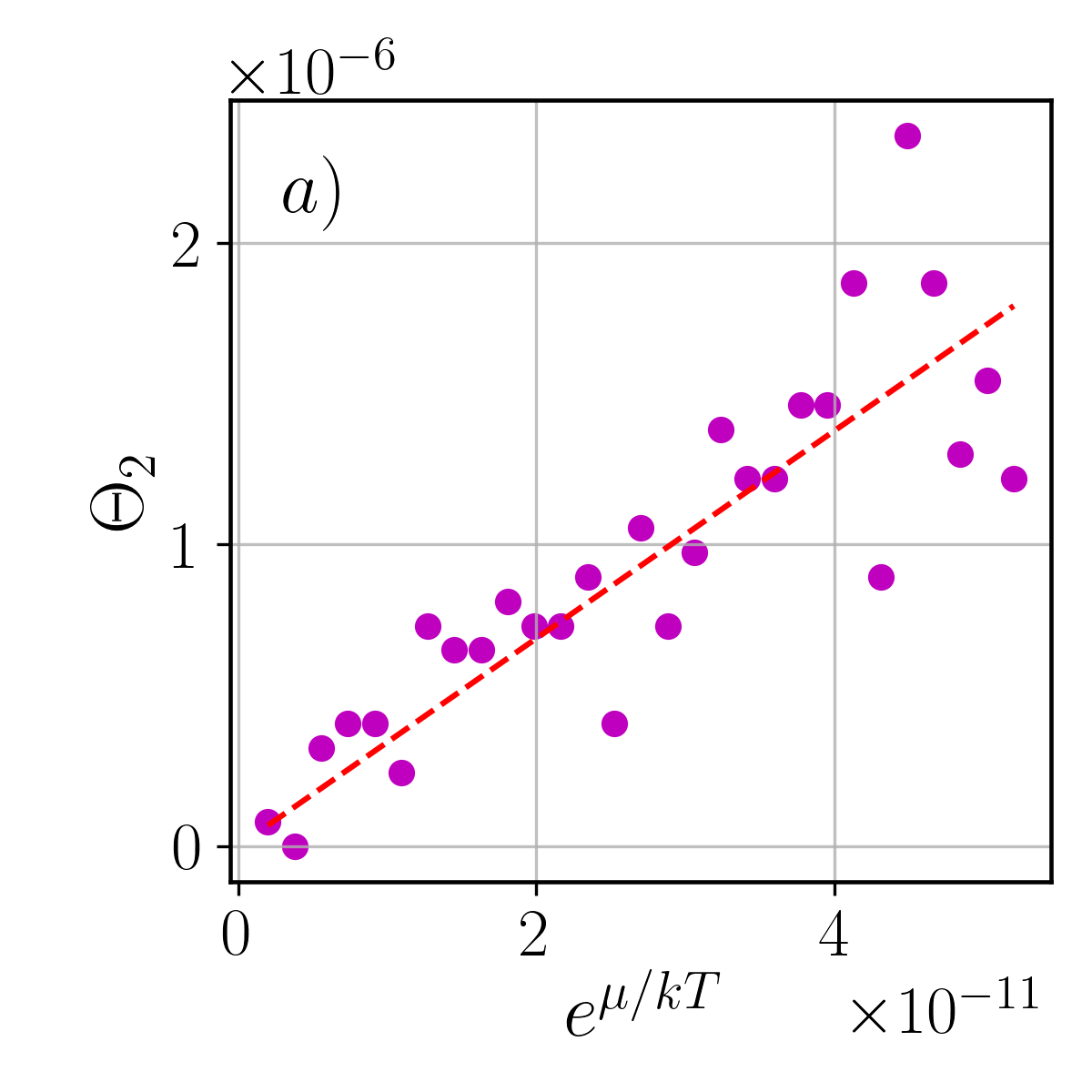} 
    \includegraphics[width=6cm]{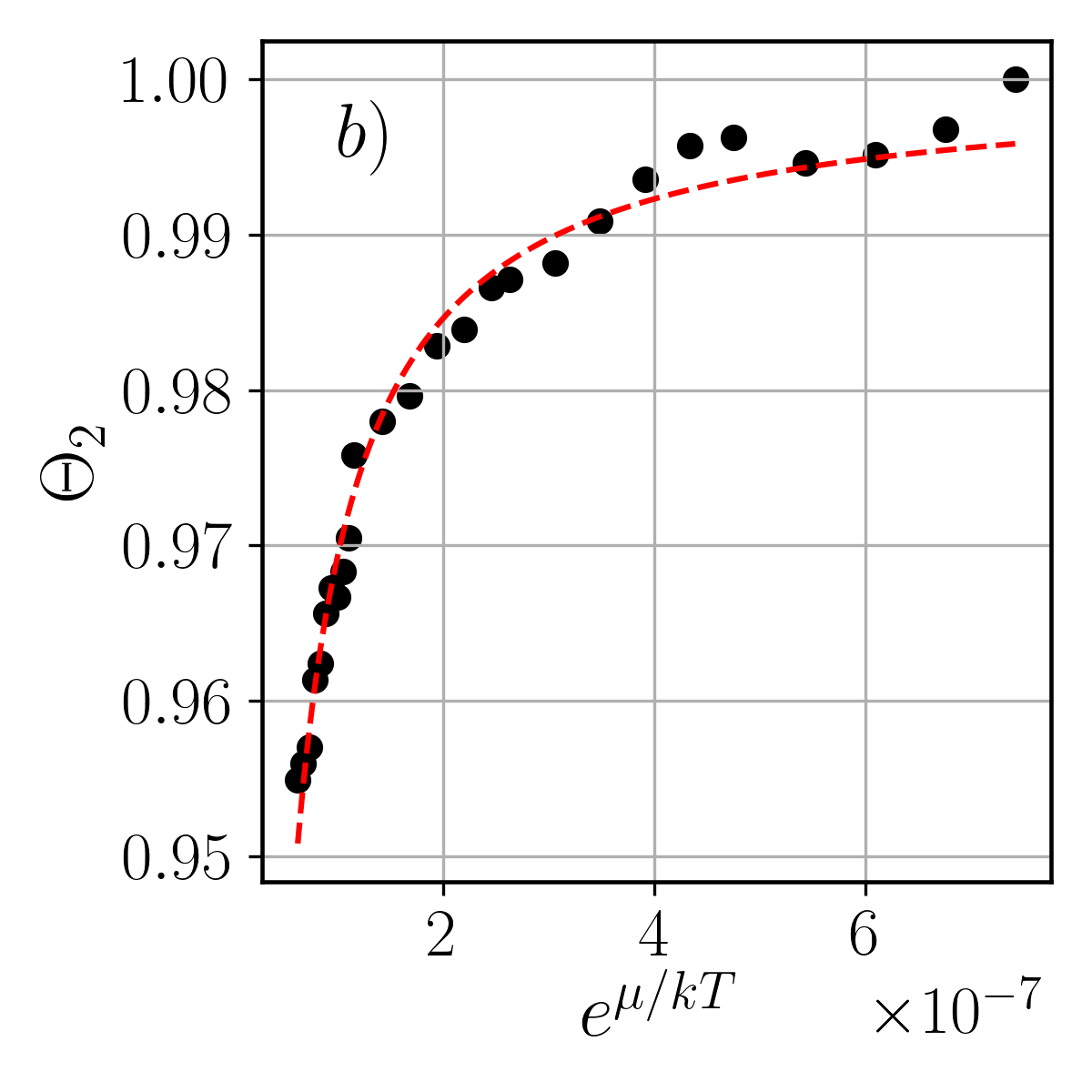} 
    % \caption{2 Figures side by side}%
     %\label{fig:example}%
     \caption{Fitting results of MD data points (scatter plots) and resulting analytical function (red dashed curves) for $a)$ $\Theta_2\to 0$ and $b)$ $\Theta_2 \to 1$ in the $L_2$ system and $T=350K$. The values obtained for $\epsilon_s$ and $\epsilon$ can be found in Tables 5 and 6, respectively.}
    \label{fig:mfcomp}
 \end{figure}

\section{Link to videos of the adsorption process}
\label{vid}
\begin{itemize}
    \item \textbf{Monolayer:} \url{https://www.youtube.com/watch?v=tu_1t-y23n0}
    \item \textbf{Bilayer:} \url{https://www.youtube.com/watch?v=tu_1t-y23n0}
     \item \textbf{Trilayer:} \url{https://www.youtube.com/watch?v=o2L6FrdHn-g}
      \item \textbf{Quadrilayer:} \url{https://www.youtube.com/watch?v=o2L6FrdHn-ge}
\end{itemize}